\documentclass[12pt]{article}

\usepackage{scicite}
\usepackage{graphicx}
\usepackage{etoolbox}
\usepackage{hyperref}
\usepackage{times}
\usepackage{bm}
\usepackage{makecell}
\usepackage{color}
\usepackage{xcolor}
\usepackage{amssymb}
\usepackage{amsmath}
\usepackage{braket}
\usepackage{siunitx}
\usepackage{lineno}
\usepackage{multicol}

\newenvironment{sciabstract}{%
\begin{quote} \bf}
{\end{quote}}

\usepackage{graphicx}

\graphicspath{ {figures/} }

\makeatletter
\let\saved@includegraphics\includegraphics
\AtBeginDocument{\let\includegraphics\saved@includegraphics}
\makeatother

\DeclareSIUnit\gauss{G}

\begin{document}

\title{Imaging the Meissner effect and flux trapping in a hydride superconductor at megabar pressures using a nanoscale quantum sensor}

\date{}


\author
{P.~Bhattacharyya,$^{1,2,\ast}$ 
W.~Chen,$^{3,\ast}$ 
X.~Huang,$^{3,\ast}$ 
S.~Chatterjee,$^{1,4}$
B.~Huang,$^{5}$\\
B.~Kobrin,$^{1,2}$
Y.~Lyu,$^{1}$ 
T.~J.~Smart,$^{1,6}$
M.~Block,$^{7}$
E.~Wang,$^{8}$
Z.~Wang,$^{7}$\\
W.~Wu,$^{7}$
S.~Hsieh,$^{1,2}$
H.~Ma,$^{5}$
S.~Mandyam,$^{7}$
B.~Chen,$^{7}$ 
E.~Davis,$^{1}$\\
Z.~M.~Geballe,$^{9}$ 
C.~Zu,$^{10}$
V.~Struzhkin,$^{11}$
R.~Jeanloz,$^{6}$ 
J.~E.~Moore,$^{1,2}$
T.~Cui,$^{3,12}$\\
G.~Galli,$^{5,13,14}$
B.~I.~Halperin,$^{7}$
C.~R.~Laumann,$^{15}$
N.~Y.~Yao$^{1,2,7\dagger}$\\
\\
\normalsize{\hspace{-8mm}$^{1}$Department of Physics, University of California, Berkeley, CA 94720, USA}\\
%
\normalsize{\hspace{-8mm}$^{2}$Materials Science Division, Lawrence Berkeley National Laboratory, Berkeley, CA 94720, USA}\\
%
\normalsize{\hspace{-8mm}$^{3}$State Key Laboratory of Superhard Materials,}\\
\normalsize{\hspace{-8mm}College of Physics, Jilin University, Changchun 130012, China}\\
%
\normalsize{\hspace{-8mm}$^{4}$ Department of Physics, Carnegie Mellon University, Pittsburgh, PA 15213, USA}\\
%
\normalsize{\hspace{-8mm}$^{5}$ Department of Chemistry, University of Chicago, Chicago, IL 60637, USA}\\
%
\normalsize{\hspace{-8mm}$^{6}$ Department of Earth and Planetary Science, University of California, Berkeley, CA 94720, USA}\\
%
\normalsize{\hspace{-8mm}$^{7}$ Department of Physics, Harvard University, Cambridge, MA 02135, USA}\\
%
\normalsize{\hspace{-8mm}$^{8}$ Department of Chemistry and Chemical Biology, Harvard University, Cambridge, MA 02135, USA}\\
%
\normalsize{\hspace{-8mm}$^{9}$ Earth and Planets Laboratory, Carnegie Institution of Washington, Washington DC 20015, USA}\\
%
\normalsize{\hspace{-8mm}$^{10}$ Department of Physics, Washington University in St. Louis, St. Louis, MO 63130, USA}\\
%
\normalsize{\hspace{-8mm}$^{11}$ Center for High Pressure Science and Technology Advanced Research, Shanghai 201203, China}\\
%
\normalsize{\hspace{-8mm}$^{12}$ School of Physical Science and Technology, Ningbo University, Ningbo 315211, China}\\
\normalsize{\hspace{-8mm}$^{13}$ Materials Science Division and Center for Molecular Engineering,}\\
\normalsize{\hspace{-8mm}Argonne National Laboratory, Lemont, IL 60439, USA}\\
%
\normalsize{\hspace{-8mm}$^{14}$ Pritzker School of Molecular Engineering, University of Chicago, Chicago, IL 60637, USA}\\
%
\normalsize{\hspace{-8mm}$^{15}$ Department of Physics, Boston University, Boston, MA 02215, USA}\\
\normalsize{$^*$These authors contributed equally to this work.}\\
\normalsize{\hspace{-8mm}$^\dagger$To whom correspondence should be addressed; E-mail: nyao@fas.harvard.edu.}
}
\date{}

\baselineskip24pt
\maketitle

\vspace{5mm}

\begin{sciabstract}
By directly altering microscopic interactions, pressure provides a powerful tuning knob for the exploration of  condensed phases and geophysical phenomena~\cite{mao2018solids}.
The megabar regime represents an exciting frontier, where recent discoveries include novel high-temperature superconductors, as well as structural and valence phase transitions~\cite{wang2021future,lilia20222021,zhang2006valence,loubeyre2020synchrotron,weck2022evidence,chen2020persistent}. 
However, at such high pressures, many conventional measurement techniques fail.
%
%
%
Here, we demonstrate the ability to perform local magnetometry inside of a diamond anvil cell with sub-micron spatial resolution at megabar pressures.
Our approach utilizes a shallow layer of Nitrogen-Vacancy (NV) color centers implanted directly within the anvil~\cite{hsieh2019imaging,lesik2019magnetic,steele2017optically}; crucially, we choose a crystal cut compatible with the intrinsic symmetries of the NV center to enable functionality at megabar pressures.
We apply our technique to characterize a recently discovered hydride superconductor, CeH\textsubscript{9}~\cite{chen2021high}.
By performing simultaneous magnetometry and electrical transport measurements, we observe the dual signatures of superconductivity: local diamagnetism characteristic of the Meissner effect and a sharp drop of the resistance to near zero.
By locally mapping the Meissner effect and flux trapping, we directly image the geometry of superconducting regions, revealing significant inhomogeneities at the micron scale. 
Our work brings quantum sensing to the megabar frontier and enables the 
closed loop optimization of superhydride materials synthesis. 
\end{sciabstract}

\maketitle

\vspace{5mm}

The recent proliferation of work on superhydride materials---hydrogen-rich compounds containing rare-earth or actinide elements---is part of a long standing search for superconductivity at room temperature~\cite{drozdov2015conventional,drozdov2015superconductivity,drozdov2019superconductivity,hong2020superconductivity,somayazulu2019evidence,kong2021superconductivity,troyan2021anomalous,snider2021synthesis,semenok2020superconductivity,zhou2020superconducting,semenok2021superconductivity,chen2021high,hong2022possible,chen2021synthesis,ma2022high,li2022superconductivity,he2022superconductivity}.
The intuition underlying this approach dates back nearly half a century~\cite{ashcroft1968metallic}: hydrogen's minimal mass and covalent bonding lead to the presence of both high-frequency phonons and strong electron-phonon interactions. 
The combination of these features is predicted to favor the formation of Cooper pairs, and thus superconductivity, at relatively high temperatures~\cite{ashcroft2004hydrogen}. 
This strategy has been fruitful, leading to the discovery and characterization of nearly a dozen superconducting hydrides in the last decade~\cite{wang2021future,lilia20222021}. 
The synthesis of these materials relies upon the application of megabar ($\sim 100$~GPa) pressures using diamond anvil cells (DACs).
This requirement naturally constrains the size and homogeneity of the samples, significantly complicating attempts at \emph{in situ} characterization. 
For example, it is extremely challenging for conventional probes to image the geometry of superconducting grains or to measure local properties.

This challenge is particularly acute for studying the magnetic signatures of superconductivity~\cite{eremets2022high,hirsch2021absence}.
%
Typical probes of magnetism average over the entire DAC geometry thereby discarding information encoded in local spatial features. 
In combination with small sample volumes, the averaging often leads to a weak magnetic signal that competes with a relatively large background. 
The ability to perform spatially-resolved magnetometry near the hydride sample would overcome these challenges and enable both enhanced field sensitivities, as well as local measurements of the Meissner effect and flux trapping.
Doing so together with resistance measurements would allow one to simultaneously probe the key electrical and magnetic signatures of superconductivity. 

\begin{figure}[ht]
  \vspace{-15mm}
 \includegraphics[width=\textwidth]{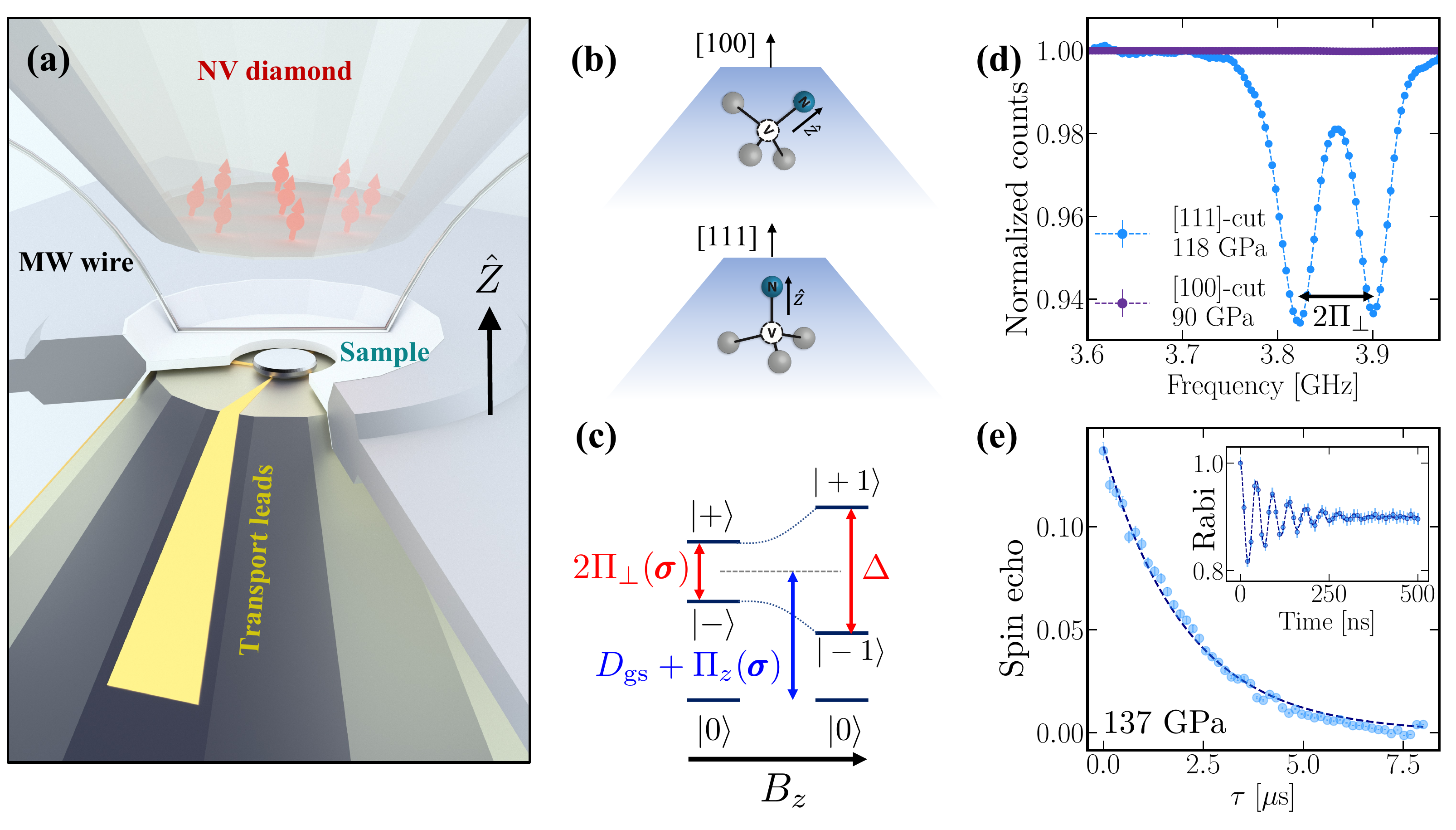}
 \centering
 \caption{\textbf{NV sensing at megabar pressures.}
 \textbf{(a)} Schematic of the sample loading showing CeH\textsubscript{9} compressed between two opposing anvils. 
 The top anvil contains a shallow layer of NV centers ($\sim1$~ppm density) approximately $50$~nm below the culet surface. 
 For ODMR measurements, a platinum wire is placed on the top culet to deliver microwaves. 
 Transport leads for four point resistance measurements are patterned on the bottom anvil. 
 An insulating gasket is used to isolate the leads and the microwave wire.
 The loading axis ($\hat Z$) is defined by the culet normal.
 \textbf{(b)}
 The quantization axis ($\hat z$) of the NV center defines its local frame.
 The crystal cut of the diamond anvil determines the projection of culet stresses in the NV frame.
 For a [100]-cut anvil (top), the dominant culet stresses ($\sigma_{ZZ}$ and $\sigma_{\perp}$) break the C\textsubscript{3v}-symmetry of all four NV subgroups.
 For a [111]-cut anvil (bottom), these stresses preserve the C\textsubscript{3v}-symmetry of the specific NV subgroup whose quantization axis is coincident with the loading axis (shown). 
 For this particular NV subgroup, we observe excellent ODMR contrast up to pressure of $\sim140$~GPa. 
 \textbf{(c)} Schematic depiction of the NV's spin sublevels in the presence of stress and magnetic field. 
  Symmetry preserving stresses, quantified by $\Pi_{z}$, directly add to the zero field splitting, $D_{\textrm{gs}}$, while symmetry breaking stresses induce a splitting,
  $2 \Pi_{\perp}$.
 An axial magnetic field $B_z$ induces a Zeeman splitting that adds in quadrature to the stress splitting.
 \textbf{(d)} 
 A continuous wave ODMR measurement on a [111]-cut anvil (sample S1) showing $\sim 6 \%$ contrast at $\approx 118$~GPa and a splitting, $2 \Pi_{\perp} \sim(2\pi)\times78$~MHz (blue data points). 
 For comparison, the ODMR contrast in a [100]-cut anvil at $\approx 90$~GPa is $\sim0.01\%$ (purple data points). 
 \textbf{(e)} A spin echo (i.e.~pulsed) measurement on sample S2 at $137$~GPa yields an NV coherence time, $T_2^{\textrm{echo}} = 2.04(4)$~\SI{}{\micro\second}.
 We demonstrate Rabi frequencies of up to $\sim(2\pi)\times25$~MHz (inset).
 }
\label{fig:1}
\end{figure}

In this work, we develop a novel platform for metrology at megabar pressures based upon the nitrogen-vacancy (NV) color center in diamond~\cite{hsieh2019imaging,lesik2019magnetic,yip2019measuring,steele2017optically}. 
By instrumenting diamond anvils with shallow ensembles of NV centers, we directly image both the Meissner effect and flux trapping with sub-micron resolution in a cerium superhydride (CeH\textsubscript{9}).
Our main results are three-fold.
First, by utilizing NVs embedded in a [111]-crystal cut anvil [Fig.~\ref{fig:1}], we demonstrate the ability to perform both DC and AC magnetometry at pressures up to $\sim 140$~GPa. 
Second, leveraging this ability, we show that CeH\textsubscript{9} locally suppresses an external magnetic field after zero field cooling. 
Upon field cooling, we observe the partial expulsion of magnetic fields below the superconducting transition temperature, $T_\textrm{c}$, as determined by simultaneous electrical resistance measurements. 
By spatially mapping these observations, we are able to directly measure the size and geometry of the superconducting regions. 
Finally, by cycling both magnetic fields and temperature, we investigate the presence of hysteresis in the magnetization.
We observe signatures of flux trapping whose strength depends on the temperature and cooling history. 

\begin{figure}[ht]
 \centering
 \includegraphics[width=\textwidth]{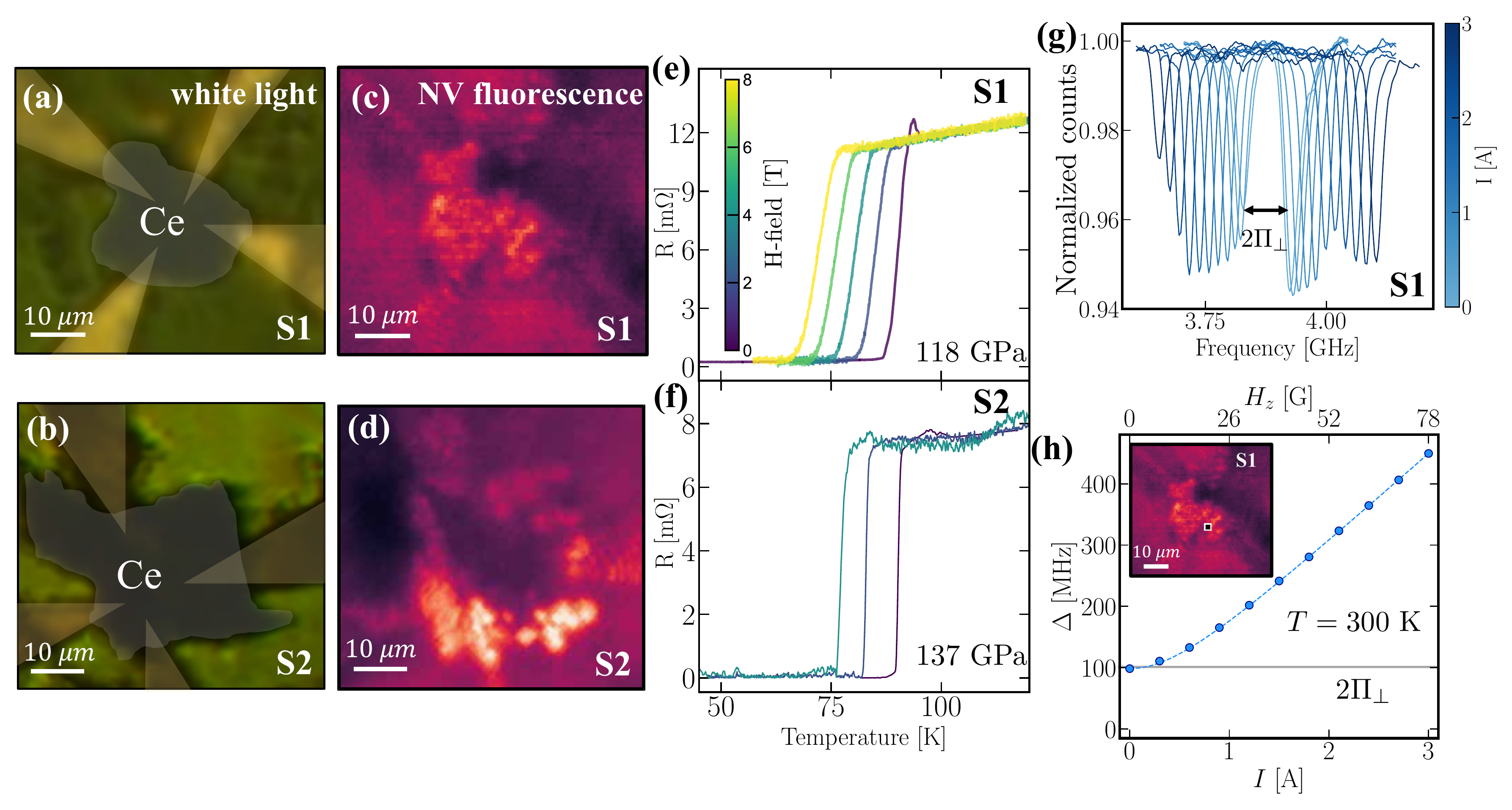}
 \caption{\small 
 \textbf{Sample synthesis and characterization.}
 To prepare samples S1 and S2, we compress a mixture of cerium (Ce) metal and ammonia borane (NH\textsubscript{3}BH\textsubscript{3}) to pressures $> 100$~GPa and laser heat ($\sim 1500$~K) to synthesize cerium hydride. 
 White light microscope images of \textbf{(a)} S1 and \textbf{(b)} S2 highlighting the Ce metal (grey) and transport leads (yellow).
 Confocal NV fluorescence maps of the corresponding regions in \textbf{(c)} S1 and \textbf{(d)} S2 also show the sample and the leads.
 Bright regions on the Ce metal in S1 and around the Ce metal in S2 correspond to additional NV centers that are created by laser heating (see Methods).
 Electrical resistance measurements of both \textbf{(e)} S1 and \textbf{(f)} S2 exhibit a sharp drop, with $T_{\textrm{c}}(P) \approx 90$~K, suggesting the formation of CeH\textsubscript{9}~\cite{chen2021high}.
 $T_{\textrm{c}}$ is suppressed on the application of magnetic fields. 
 \textbf{(g-h)} Calibration of the applied field, $H_z$, at room temperature ($T=300$~K).
We generate the field, $H_z$, by applying a current, $I$, in an electromagnet.
 Starting with a stress induced splitting $2 \Pi_{\perp}$ at $I=0$, we measure an increase in the ODMR splitting, $\Delta$, with increasing current.
 \textbf{(g)} ODMR spectra measured for different values of $I$ at a specific spatial point on sample S1 [inset of (h)].
 The ODMR splitting $\Delta = \sqrt{(2 \Pi_{\perp})^2 + (2\gamma_B B_z)^2}$ is a quadrature sum of the stress splitting, $2 \Pi_{\perp} \sim(2\pi)\times98$~MHz, and the Zeeman splitting, $2\gamma_B B_z$. 
 In the absence of sample magnetism (i.e., for $T > T_{\textrm{c}}$), $B_z = H_z \propto I$. 
 \textbf{(h)} Fitting the measured splitting $\Delta$ to this functional form, we directly extract $H_z$ (in gauss), calibrated for each value of $I$ (in amperes). 
 We verify that this calibration is temperature independent (see Methods).
 }\label{fig:2}
\end{figure}

Our experiments are performed on two independent samples (S1 and S2) of cerium superhydride prepared via laser-heating inside miniature panoramic diamond anvil cells~\cite{gavriliuk2009miniature} [Fig.~\ref{fig:2}(a-d)] (see Methods). 
Four point electrical transport measurements [Fig.~\ref{fig:2}(e-f)] on both samples exhibit a sharp drop in resistance at $T_\textrm{c} \approx 90$~K, suggesting the formation of CeH\textsubscript{9}~\cite{chen2021high}. 
For each DAC, the top anvil is a Type Ib [111]-cut anvil, which is implanted with a layer of NV centers $\sim 50$~nm below the culet surface at a density of approximately $\sim 1$~ppm~\cite{hsieh2019imaging,lesik2019magnetic}.
Each NV center hosts a spin-1 electronic ground state governed by the Hamiltonian, $H_0 = D_{\textrm{gs}} S_z^2$.
Here, $\vec{S}$ are the NV's spin-1 operators 
with the N-V axis defining the quantization axis ($\hat z$) 
and $D_{\textrm{gs}} = (2\pi)\times \SI{2.87}{\giga\hertz}$ being the zero-field splitting between the $|m_\textrm{s}= 0\rangle$ spin sub-level and the degenerate $|m_\textrm{s}= \pm 1\rangle$ sub-levels~\cite{doherty2013nitrogen}.
Local perturbations such as temperature, stress, electric, and magnetic fields couple to the NV center and change the energy of its spin states~\cite{acosta2010temperature,maze2008nanoscale,dolde2011electric,ovartchaiyapong2014dynamic,doherty2014electronic,barson2017nanomechanical}.
These changes can be read-out via optically detected magnetic resonance (ODMR) spectroscopy where one measures a change in the NV's fluorescence upon chirping a microwave field through resonance (see Methods)~\cite{schirhagl2014nitrogen}. 

Our central explorations of CeH\textsubscript{9} involve local measurements (at $\gtrsim 100$~GPa) of the magnetic field at the NV center, $B_z$, as one tunes the temperature, $T$, and an external magnetic field, $H_z$, applied along the NV axis, $\hat{z}$.
Thus, it is crucial to be able to separate out the effects of crystal stress and temperature from the NV's ODMR spectrum, in order to accurately measure $B_z$.
Luckily, this is a relatively straightforward task. 
At megabar pressures, the effects of stress dominate over those of temperature by nearly three orders of magnitude (see Methods), so we focus on deconvolving the effects of the local stress tensor, $\pmb{\sigma}$.
Stress couples to the NV center via the effective Hamiltonian, $H_\textrm{S} = \Pi_z S_z^2 + \Pi_x (S_y^2 - S_x^2) + \Pi_y (S_x S_y + S_y S_x)$, where the parameters $\Pi_i= \Pi_i(\pmb{\sigma})$ depend on the appropriate components of $\pmb{\sigma}$  (see Methods).
$\Pi_{z}$ captures the additional zero-field splitting due to C\textsubscript{3v}-symmetry-preserving stresses [Fig.~\ref{fig:1}(c)]~\cite{barson2017nanomechanical,hsieh2019imaging}.
Meanwhile, $\Pi_{\perp}=\sqrt{\Pi_x^2 +\Pi_y^2}$ parametrizes the symmetry-breaking stresses, mixing the $\ket{m_{\textrm{s}}=\pm 1}$ spin states into new eigenstates, $\ket{\pm}=\left(\ket{m_{\textrm{s}}=+1} \pm e
^{i\phi_\Pi} \ket{m_{\textrm{s}}= -1}\right)/\sqrt{2}$, which are split by $2 \Pi_{\perp}$ [Fig.~\ref{fig:1}(c,d)]; here, $\phi_\Pi=\arctan(\Pi_y/\Pi_x)$. 
Since the Zeeman splitting from an axial magnetic field adds in quadrature with the stress splitting, the NV's ODMR spectrum exhibits a pair of resonances with a total splitting of $\Delta = \sqrt{(2 \Pi_{\perp})^2 + (2\gamma_B B_z)^2}$, where $\gamma_B = (2\pi)\times\SI{2.8}{\mega\hertz/\gauss}$ is the NV's gyromagnetic ratio 
[Fig.~\ref{fig:1}(c)].
This provides a simple prescription for measuring $B_z$: 
extract $\Delta$ from the NV's ODMR spectrum and then subtract (in quadrature) the stress-induced splitting, $2 \Pi_{\perp}$, measured at zero field. 
 
While conceptually simple, extending this approach to megabar pressures~\cite{dai2022optically,hilberer2023nv} has been riddled with persistent challenges.
These include: diminishing NV fluorescence, significant broadening of the ODMR spectrum, and a dramatic loss of optical contrast above $\sim50$~GPa~\cite{doherty2014electronic}.
We address these difficulties by introducing a new approach to NV-based metrology at megabar pressures.
In particular, by using a [111]-cut anvil, we engineer the dominant culet stress to project along the quantization axis for one specific subgroup of NV centers [Fig.~\ref{fig:1}(b)]. 
Doing so has been conjectured to reduce the loss of contrast at high pressures~\cite{goldman2015state,davies1976optical,bencheninprep} (see Methods).
%
This is indeed borne out by the data shown in Fig.~\ref{fig:1}(d) --- the ODMR contrast for NV's in a [111]-cut anvil is two orders of magnitude larger than that in a [100]-cut anvil. 
This enables us to characterize both continuous wave (CW) and pulsed measurements up to pressures of order $\sim 140$~GPa.
For CW measurements, we achieve a maximum of $\sim 15 \%$ contrast at megabar pressures, roughly an order of magnitude larger than that obtained under quasi-hydrostatic stresses. This leads to typical magnetic field sensitivities of $\sim~35~\SI{}{\micro\tesla/\sqrt{\hertz}}$ (see Methods).
For pulsed measurements, we perform spin echo and measure a coherence time, $T_2^{\textrm{echo}} = 2.04(4)$~\SI{}{\micro\second} at $137$~GPa [Fig.~\ref{fig:1}(e)].

Let us now turn to calibrating the external magnetic field at megabar pressure. 
The field strength is tuned via the current, $I$, applied to an electromagnet; Fig.~\ref{fig:2}(g) depicts the ODMR spectra of sample S1 at room temperature ($T=300$~K) as the applied current is increased. 
By extracting the ODMR splitting $\Delta$, at each value of the current, and then subtracting the $2 \Pi_{\perp}$ splitting measured at $I=0$, one can immediately convert the applied current to an external magnetic field strength, $H_z$ [Fig.~\ref{fig:2}(h)].
This conversion assumes that the CeH$_9$ sample does not contribute an appreciable additional field above the superconducting transition temperature and thus, that $B_z = H_z$ (in Gaussian units).
We verify the robustness of this assumption by confirming that $H_z$ is independent of both the temperature and the spatial location within the sample chamber (see Methods). 
Furthermore, we note that sample S2 yields the same current-to-field calibration. 

\begin{figure}[ht]
 \hspace{-3mm}
 \includegraphics[width=\textwidth]{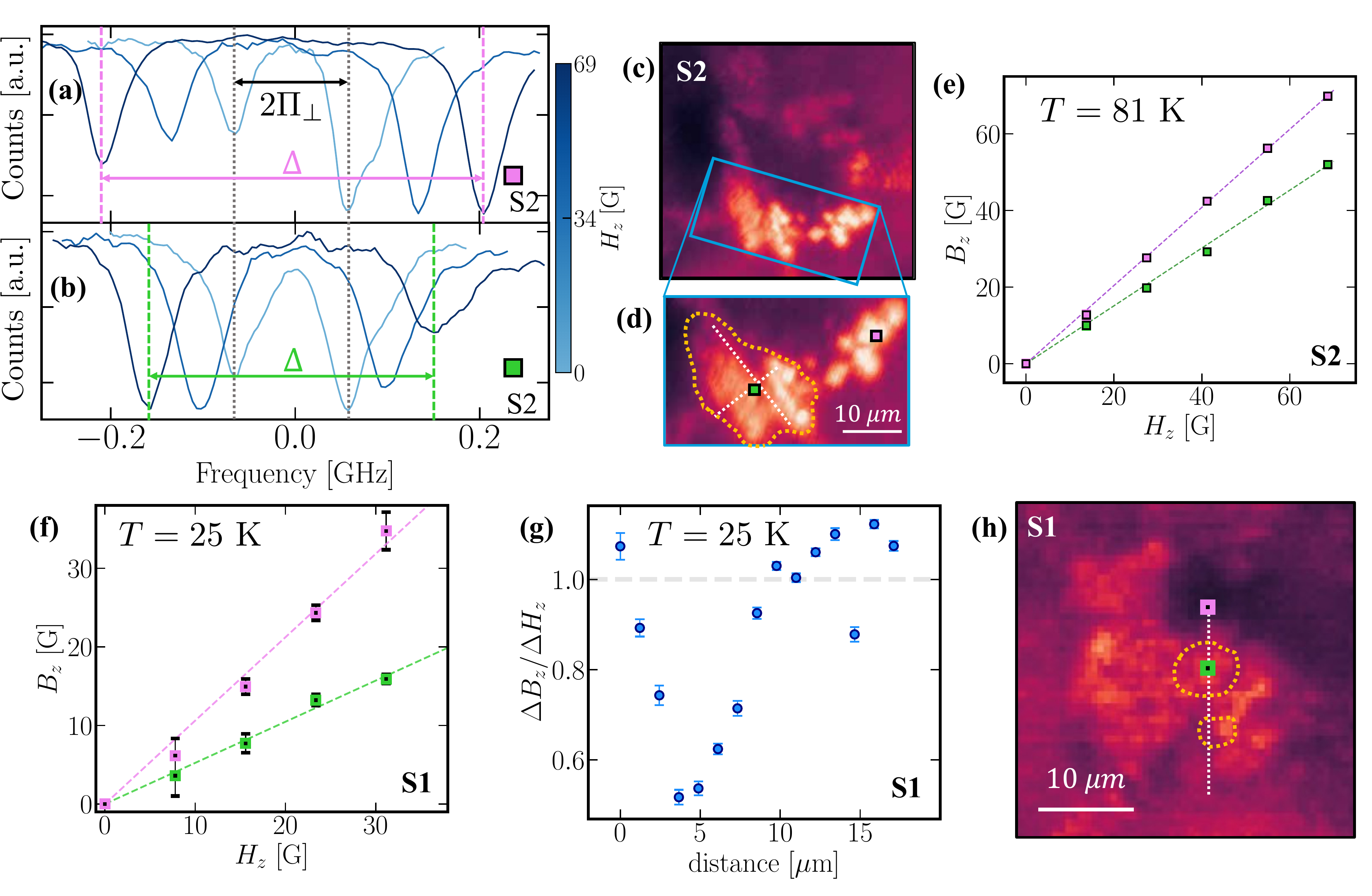}
 \caption{\small \textbf{Spatially resolved measurements of the Meissner effect in CeH\textsubscript{9}.}
\textbf{(a-b)} ODMR spectra collected on zero field cooling to a temperature $T<T_{\textrm{c}}$, at two representative spatial points in sample S2 [shown in (d)].
The purple point (a) is far from the CeH\textsubscript{9} sample, while the green point (b) is directly above it. 
 The $2 \Pi_{\perp}$ stress splittings measured at $H_z = 0$ are similar at both locations.
 However, as the applied external field, $H_z$, is increased (at $T = 81~\textrm{K}$), the ODMR splitting 
 above the CeH\textsubscript{9} (b) is suppressed relative to the splitting away from it (a), reflecting a local diamagnetic response.
 For clarity, all spectra have been centered by subtracting the ODMR shift.
 In addition, the peak contrast has been rescaled to clearly show the differences in splitting.
 \textbf{(c)} Confocal fluorescence image of S2. 
 \textbf{(d)} 
By performing ODMR at different spatial points, we identify a sub-region of extent $\sim10$~\SI{}{\micro\meter} exhibiting local diamagnetism [delineated by the dotted yellow line in (d)].
We in fact use this signal to identify the regions where CeH\textsubscript{9} has been successfully synthesized via laser heating. 
 \textbf{(e)} For the purple (green) point sitting outside (inside) this sub-region, we plot $B_z$ extracted from the ODMR spectra (a-b) at each value of the applied external field, $H_z$.
 Away from the CeH\textsubscript{9} (purple), we measure a slope $s= \Delta B_z/\Delta H_z = 1.02$, suggesting that there is no local magnetization. 
 In contrast, above the CeH\textsubscript{9} (green), we measure a slope $s= 0.75$, demonstrating a clear local suppression of the external magnetic field ($B_z < H_z$), consistent with the Meissner effect. 
 \textbf{(f)} We perform analogous measurements at $T= 25$~K in sample S1 [green and purple points in (h)].
 In comparison to S2, we observe a stronger suppression ($s= 0.52$) above the CeH\textsubscript{9}.
 %
 \textbf{(g)} By measuring the slope $s= \Delta B_z / \Delta H_z$ at $T=25$~K along a line cut [dotted white line in (h)] we image the spatial profile of the Meissner-induced magnetic field suppression in sample S1.
 \textbf{(h)} We identify two disconnected sub-regions of extent $< 10$~\SI{}{\micro\meter} exhibiting local diamagnetism (i.e.~$s<1$).
 }\label{fig:3}
\end{figure}

\emph{Local suppression of an external magnetic field}---In order to probe the sample's magnetic response below $T_\textrm{c}$, we perform NV magnetometry in a cryogenic system (down to $T=25$~K) integrated with a scanning confocal microscope~(see Methods).
Beginning with sample S2 at $300$~K, we zero-field cool (at $P=137$~GPa) below the transition temperature ($T_\textrm c\approx91$~K) to $T=81$~K, and perform ODMR spectroscopy at multiple points within the sample chamber. 
We focus on two representative spatial points: one above the CeH\textsubscript{9} sample [green point, Fig.~\ref{fig:3}(d)], and one far from the sample [purple point, Fig.~\ref{fig:3}(d)]. 
Starting with the distant point, at $H_z=0$, we find that the NV exhibits a $2 \Pi_{\perp}$ splitting of $\sim(2\pi)\times134$~MHz [light blue curve, Fig.~\ref{fig:3}(a)]. 
As expected, turning on the external magnetic field (up to $H_z = 70$~G) causes the spectrum to further split [darker blue curves, Fig.~\ref{fig:3}(a)].
%
We extract the magnetic field, $B_z$, at the NV's location as a function of the external applied field, $H_z$, and find that $B_z = H_z$ nearly perfectly [purple data, Fig.~\ref{fig:3}(e)]. This implies that away from the CeH\textsubscript{9} sample, there is no local magnetization.

The response of NV's above the CeH\textsubscript{9} is markedly distinct.
The ODMR splitting, $\Delta$, exhibits a significantly weaker increase as the external field is ramped up [Fig.~\ref{fig:3}(b)], indicative of a local suppression in $B_z$ [green points, Fig.~\ref{fig:3}(e)].
This response is consistent with diamagnetism from the sample as would be expected from a superconducting Meissner effect. 
A similar local suppression is observed in sample S1 [Fig.~\ref{fig:3}(f)].

Our ability to locally image the Meissner effect enables us to directly characterize the geometry and size of the superconducting regions within the sample chamber. 
As an example, Fig.~\ref{fig:3}(g) depicts a particular line cut in sample S1.
For each point along the line cut, we measure $B_z$ as a function of $H_z$ and extract the slope, $s = \Delta B_z / \Delta H_z$, where $s<1$ indicates suppression of the local field. 
As shown in Fig.~\ref{fig:3}(g), there exists a contiguous region where $s<1$, suggesting the presence of a $\lesssim 10\mu$m sized region of superconducting CeH\textsubscript{9}.
Similar sized regions of CeH\textsubscript{9} are also observed in sample S2 (see e.g. the orthogonal line cuts indicated in Fig.~\ref{fig:3}(d) and the Methods). 
Our spatial surveys point to an intriguing observation: although synthesis is performed by rastering a high-power laser, only a  fraction of the laser-heated area exhibits superconductivity.
%
%

\begin{figure}
 \vspace{-20mm}
  \hspace{-10mm}
 \includegraphics[width=1.1\textwidth]{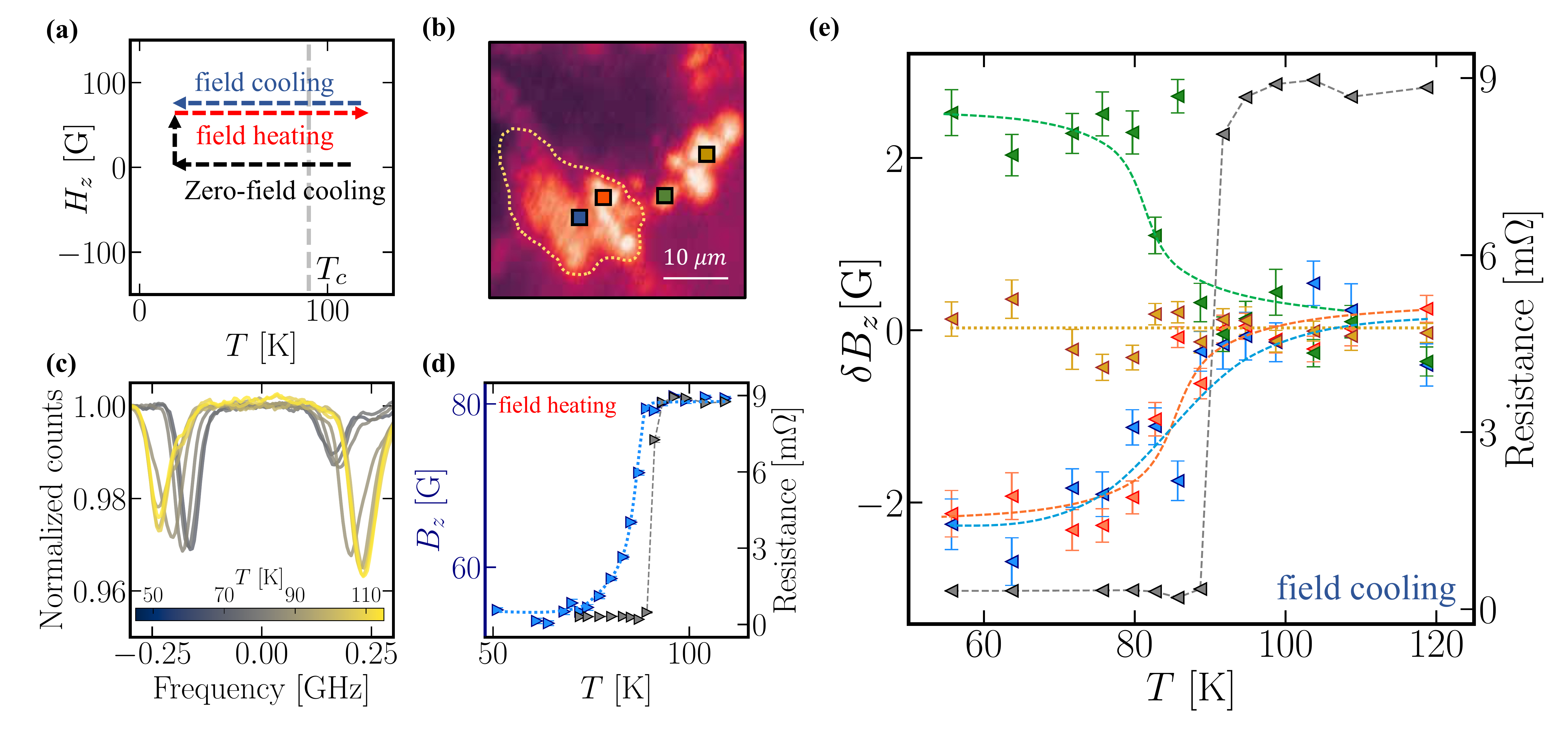}
 \vspace{-8mm}
 \caption{\small \textbf{Simultaneous measurements of electrical resistance and magnetism on sweeping $T$.}
 \textbf{(a)} Schematic depicting experimental protocols.
 Zero field cooling (black): sample is cooled below $T_\textrm{c}$ with zero applied field, $H_z = 0$, to a specific temperature, and then an external field is applied. 
 Field heating (red): sample is heated above $T_\textrm{c}$ with $H_z$ held constant.
 Field cooling (blue): sample is cooled below $T_\textrm{c}$ in the presence of a finite applied field $H_z$.
 The temperature values shown include a correction for thermal gradients in our cryostat (see Methods).
 \textbf{(b)} Confocal fluorescence image of S2 zoomed into the same CeH\textsubscript{9} region as Fig.~\ref{fig:3}(d).
 NV ODMR spectroscopy is performed at four spatial points: two points above the CeH\textsubscript{9} region (blue and red), one point at the edge of this region (green), and one point far away from this region (yellow).
 \textbf{(c)} NV ODMR spectra collected at the blue spatial point in (b) on field heating at $H_z=79$~G (following zero field cooling).
 The ODMR splitting $\Delta$ increases as $T$ is increased across $T_\textrm c$. 
 \textbf{(d)} Shows the local field, $B_z$, (left y-axis) extracted from the ODMR splitting $\Delta$ shown in (c). The four point resistance (right y-axis) is measured simultaneously.
 We observe a sharp transition in the resistance at $T_\textrm c \approx91$~K. 
 The local magnetic field, $B_z$, exhibits a low temperature plateau at $B_z\sim53$~G~$<H_z$ for $T<72$~K (deep in the superconducting phase); $B_z$ gradually increases for $73$~K$<T<90$~K.
 Finally, concurrent with the transition measured in the resistance, we observe a second plateau in $B_z$ whose value matches that of the external applied field $H_z$.
 \textbf{(e)} Simultaneous measurements of four point resistance (right y-axis) and the change in the local field, $\delta B_z$, (left y-axis) at the four spatial points in (b) on field cooling with $H_z=79$~G. 
 The measured resistance identifies a clear transition, at $T_\textrm c \approx91$~K.
 At each spatial point, we determine $\delta B_z$ relative to the average value of the measured $B_z$ in the normal state (i.e., for temperature $T>91$~K).
 For the point far away from the CeH\textsubscript{9} region (yellow), there is no change in the local $B_z$ across the superconducting transition.
 For the points above the CeH\textsubscript{9} region (red and blue), $B_z$ decreases by $\sim2$~G across the  transition; this observation is consistent with magnetic flux expulsion as would be expected from a superconductor. 
 Interestingly, the magnitude of the change in $B_z$ is significantly smaller than the field suppression observed in (d), suggesting that magnetic flux is only partially expelled under field cooling conditions. 
 Finally, for the point at the edge of the CeH\textsubscript{9} region (green), $B_z$ increases by $\sim2~$G across the  transition.
 This is qualitatively consistent with the rearrangement of the magnetic flux due to partial expulsion by the nearby CeH$_9$ sample.
 }
  \label{fig:4}
\end{figure}

\emph{Simultaneous magnetometry and resistance measurements}---Focusing on sample S2, we now characterize the local field suppression as a function of increasing temperature.
In particular, we follow the experimental sequence depicted in Fig.~\ref{fig:4}(a) (black and red curves): We begin by zero-field cooling the sample below $T_\textrm{c}$, then we ramp up an external magnetic field, and finally, fixing this field ($H_z=79$~G), we slowly increase the temperature above $T_\textrm{c}$. 
During the field heating sequence, we measure both four-terminal resistance as well as the NV's ODMR spectrum [Fig.~\ref{fig:4}(d)].
The resistance exhibits a clear jump at $T_\textrm{c} \approx 91$~K. 
The behavior of the local magnetic field measured by the NV centers is more subtle [Fig.~\ref{fig:4}(c-d)]~\cite{nusran2018spatially,lesik2019magnetic}. 
In particular, for temperatures $T < 72$~K, the local field $B_z$ exhibits a plateau at $\sim 53$~G, significantly below the value of the external applied field, $H_z = 79$~G. 
For intermediate temperatures, $73$~K~$<T<90$~K, $B_z$ exhibits a slow increase, suggesting a gradual weakening of diamagnetism.
Finally, for temperatures $T>91$~K, coincident with the superconducting transition measured via resistance, $B_z$ exhibits a second plateau in agreement with the strength of the external magnetic field.
%

\emph{Partial field expulsion on field cooling}---While we have observed clear signatures of diamagnetism after zero field cooling, a complementary signature of a superconductor is the ability to expel magnetic flux upon field cooling [blue arrow in Fig.~\ref{fig:4}(a)]~\cite{tinkham2004introduction}.
We simultaneously measure resistance and perform ODMR spectroscopy (during field cooling) at four spatial points on sample S2 [Fig.~\ref{fig:4}(b)]: a pair of points (red and blue) above the identified CeH\textsubscript{9} region [enclosed by the dotted yellow line], a point (green) just outside the CeH\textsubscript{9} region, and a point (yellow) far away from the CeH\textsubscript{9} region. 
Again, the resistance exhibits a sharp transition at $\sim91$~K [grey triangles in Fig.~\ref{fig:4}(e), right y-axis].
At each spatial point, we determine the change in the local field, $\delta B_z$, across this transition relative to the average $B_z$ measured in the normal state ($T>91$~K) [Fig.~\ref{fig:4}(e), left y-axis].
Far from the CeH\textsubscript{9} region (yellow), the local field is temperature independent across the transition.
Above the CeH\textsubscript{9} region (red and blue), the local field decreases by $\sim2$~G on cooling below the transition. 
Intriguingly, near the edge of the CeH\textsubscript{9} region (green), the local field, $B_z$, increases by $\sim2$~G. 
Taken together, these observations are consistent with partial flux expulsion: as the sample is cooled below $T_\textrm{c}$, it expels magnetic flux from the CeH\textsubscript{9} region, leading to a reduction of magnetic flux directly above the sample and a concentration at the edge.

\begin{figure}[ht]
 \centering
  \vspace{-35mm}
 \includegraphics[width=\textwidth]{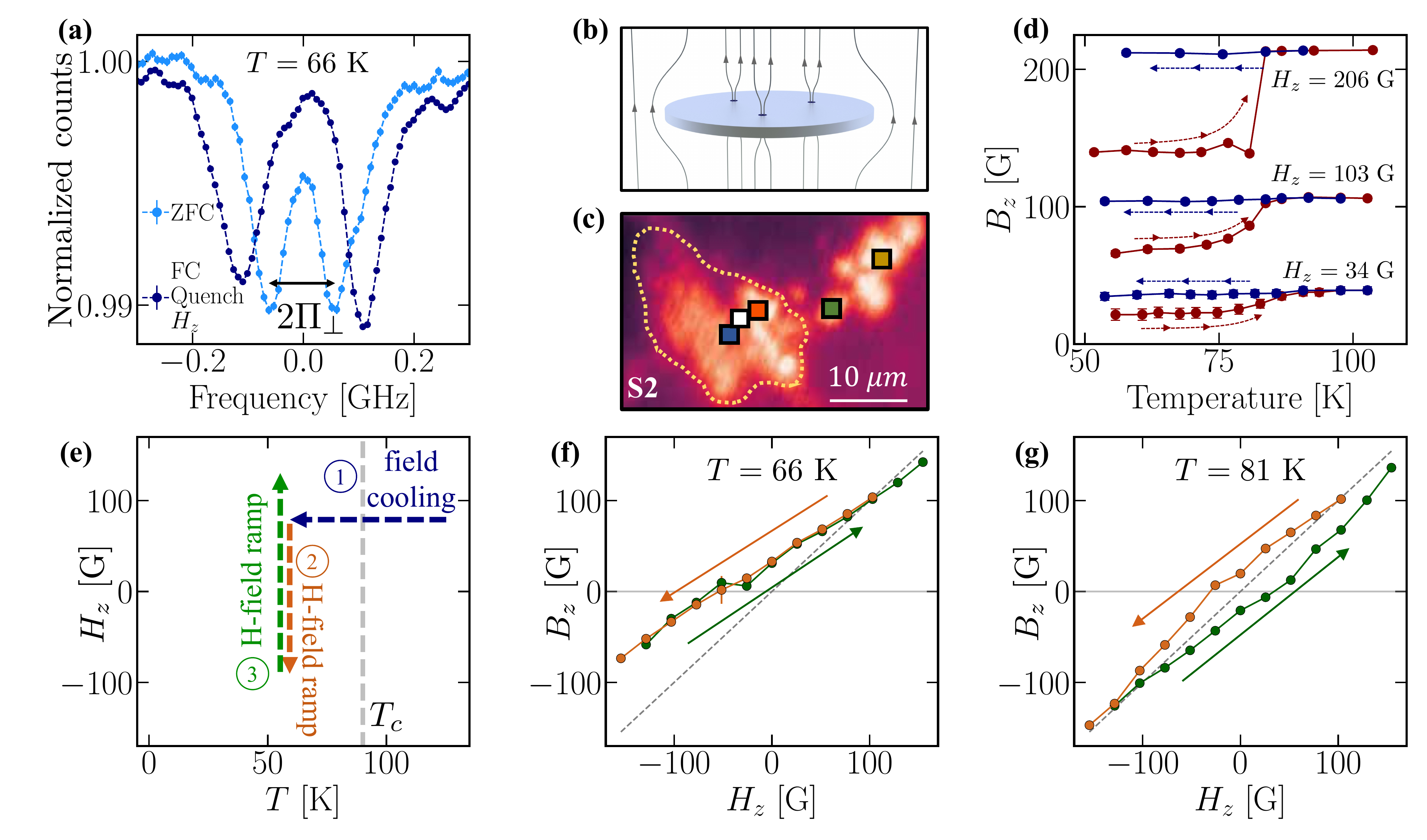}
  \vspace{-5mm}
  \caption{\small \textbf{Flux trapping and hysteresis in CeH\textsubscript{9}}.
 \textbf{(a)} A comparison of ODMR spectra at $H_z=0$, which are collected at the white spatial point in (c).
 The light blue data are obtained after zero-field cooling, while the dark blue data are measured after field cooling (at $H_z=103$~G) and subsequently quenching the external field.
The latter exhibits a splitting that corresponds to a local $B_z\sim34$~G despite the absence of any external field.
 \textbf{(b)} A schematic showing the penetration of external magnetic flux through the superconducting sample.
 For both Type I and Type II superconductors, the presence of macroscopic defects (islands of non-superconducting material) created due to incomplete synthesis may trap fluxes on field cooling.
 Additionally, for Type II superconductors, the flux may penetrate in the form of vortices when the sample is in the mixed phase~\cite{tinkham2004introduction}.
%
 %
 \textbf{(c)} Confocal fluorescence image of S2 zoomed into the same CeH\textsubscript{9} region as Fig.~\ref{fig:3}(d). All data in this figure are collected at the white spatial point above the CeH\textsubscript{9} region. 
 \textbf{(d)} Depicts measurements of $B_z$
 following the full experimental sequence shown in Fig.~\ref{fig:4}(a): zero-field cool, ramp $H_z$, field heat and then field cool. 
 The three values of $H_z$ shown correspond to the magnitude of the applied field at the end of the ramp step. 
 %
%
As $H_z$ increases, we observe a slight suppression in the transition point and an increase in the measured hysteresis (i.e.~difference in $B_z$ on field heating and cooling).
 Perhaps most notably, the transition dramatically sharpens for larger $H_z$.
 %
 \textbf{(e-g)} Measurements of hysteresis on sweeping $H_z$ at fixed $T$. 
 (e) Depicts the experimental sequence. The sample is field cooled (at $H_z=+103$~G) below the transition.
 Then, fixing $T$, we characterize the change in $B_z$ as $H_z$ is swept in magnitude and direction: first, down to $H_z= -154$~G (brown curve and data points), and then back up to $H_z=+154$~G (green curve and data points).
 In the absence of any local magnetization, one expects $B_z = H_z$ [dotted grey line in (f,g)]. 
 (f) Fixing $T=66$~K (deep in the superconducting phase), we measure a flux-trapped field ($B_z=+34$~G at $H_z=0$~G) and no appreciable hysteresis on sweeping $H_z$.
 The measured slope $s=\Delta B_z/\Delta H_z\approx0.67<1$ agrees perfectly with
 that measured after zero-field cooling for this same spatial point (see Methods).
 This suggests that, despite the presence of trapped flux, the CeH\textsubscript{9} region continues to expel additional applied fields. 
 (g) In contrast, fixing $T=81$~K (near $T_\textrm c$) and sweeping $H_z$, we observe clear hysteresis.
 In this case, we find a smaller flux-trapped field ($B_z=+20$~G at $H_z=0$~G).
 On ramping to $H_z \sim -100$~G, the 
 measured $B_z$ is nearly the same as the applied field $H_z$, without any observed change in the electrical resistance.
 Upon ramping back up to $H_z =0$,
 we observe a flux-trapped field in the opposite direction ($B_z=-23$~G).
 Assuming the flux-trapped field originates from vortices pinned within CeH\textsubscript{9}~\cite{tinkham2004introduction,matsushita2007flux}, we find that an applied field of $H_z=-154$~G is not strong enough to de-pin the vortices at $T=66$~K but is at $T=81$~K, leading to the observed hysteresis.
 }\label{fig:5}
\end{figure}

We note that the qualitative profiles of $B_z$ above the CeH\textsubscript{9} region upon field heating [Fig.~\ref{fig:4}(d)] and field cooling [Fig.~\ref{fig:4}(e)] are remarkably similar.
However, the quantitative values are quite distinct: the strength of the local field expulsion (upon field cooling) is an order of magnitude weaker than the strength of the local field suppression (after zero field cooling).
This suggests that during field cooling, the external magnetic field is able to partially penetrate through the sample [Fig.~\ref{fig:5}(b)], an observation consistent with prior measurements on superhydrides~\cite{eremets2022high}.
Associated with the penetration of magnetic flux on field cooling is the possibility of flux trapping---a textbook signature of disordered superconductors where the disorder ``pins'' the permeating magnetic field (see Methods).
This pinning leads to a remnant magnetic field, arising from frozen-in magnetic moments within the superconductor, even when the applied external field is quenched ($H_z \rightarrow 0$~G) after field cooling~\cite{minkov2206trapped,matsushita2007flux}.

\emph{Flux trapping and hysteresis of the Meissner effect}---To investigate the presence of flux trapping, we examine three spatial points above the CeH\textsubscript{9} region in sample S2 [blue, white and red points in Fig.~\ref{fig:5}(c)].
After field cooling the sample at $H_z =103$~G, we quench the external magnetic field to zero. 
The resulting ODMR spectra for one of the three spatial points [white point in Fig.~\ref{fig:5}(c)], is shown in Fig.~\ref{fig:5}(a) [dark blue curve].
By comparing to the zero-field cooled spectrum at the same spatial point [light blue curve, Fig.~\ref{fig:5}(a)], one finds that 
the ODMR splitting, $\Delta$, shows the presence of a remnant $\sim34$~G field, despite the fact that the external field, $H_z=0$~G. 
This is precisely the expected signature of flux trapping.
The same signature is observed at both of the other spatial points (see Methods).

To further explore this flux trapping, we follow the experimental sequence depicted in Fig.~\ref{fig:5}(e); in particular, after field cooling at $H_z=103$~G to a fixed temperature $T<T_\textrm c$, we ramp the external field down to $H_z=-154$~G, and then back up to $H_z=+154$~G.
For a temperature ($T=66$~K) well below the transition temperature ($T_\textrm c\approx91$~K), as the external field is ramped down, $B_z$ decreases but reaches a finite (flux-trapped) value of $\sim 34$~G at $H_z =0$~G.
As the external field switches direction ($H_z<0$), $B_z$ continues to decrease, reaching zero at $H_z \sim -51$~G; this is consistent with competition between the external magnetic field and the flux-trapped field from CeH\textsubscript{9}.
At even larger magnitudes of negative $H_z$, $B_z$ also becomes negative and scales with $H_z$.
%
Finally, as we ramp back up to positive $H_z$ fields, we observe no hysteresis in the measured $B_z$ [Fig.~\ref{fig:5}(f)].

A few remarks are in order. 
First, as a benchmark, we perform the same set of experiments [Fig.~\ref{fig:5}(e)] for a spatial point outside the CeH\textsubscript{9} region, and always observe $B_z = H_z$.
Second, the data in Fig.~\ref{fig:5}(f) exhibit a slope, $s \approx 0.67$, indicating the presence of local field suppression; interestingly, this slope is in quantitative agreement with that obtained upon applying a magnetic field after zero-field cooling (at the same spatial location, see Methods).
This suggests that in addition to the trapped flux, the Meissner effect observed on zero field cooling is also at play.
This suppression also naturally helps to explain the following observation: Although we detect a remnant flux-trapped field of $\sim34$~G, we find that this field is only canceled (i.e.~the NVs measure $B_z =0$~G) for an applied field $H_z\sim-51$~G.
%
%
Finally, we note that the data in Fig.~\ref{fig:5}(f) are taken over the course of several days. In combination with the lack of any observed hysteresis, this suggests that the flux-trapped field arises from persistent currents within the CeH\textsubscript{9} sample. 
%



On field cooling to a temperature ($T = 81$~K) near the transition, we observe distinct behavior in the measured $B_z$ as a function of magnetic field sweeps [Fig.~\ref{fig:5}(g)].
As the external field is initially ramped down, we observe analogous signatures of flux trapping with $B_z = 20$~G at $H_z = 0$~G.
However, as we switch the direction of the external field ($H_z<0$), the slope of the response changes and we find a scaling consistent with $B_z = H_z$ at the largest negative $H_z$ fields. 
On ramping back up to positive $H_z$ fields, we observe clear hysteresis in the data with signatures of flux trapping in the opposite direction, i.e.~$B_z = -23$~G at $H_z = 0$~G.
Again, as we switch the direction of the external field (back to $H_z>0$), the slope of the response changes and we find a scaling consistent with $B_z = H_z$ at the largest positive $H_z$ fields. 
In combination, Fig.~\ref{fig:5}(f) and \ref{fig:5}(g) suggest that the strength of the flux trapping is temperature dependent~\cite{matsushita2007flux}.
Moreover, as detailed in the Methods (Fig.~E11), the trapping strength itself exhibits hysteresis: on zero-field cooling to the same temperature, a significantly larger $H_z$ is required before one measures a scaling consistent with $B_z = H_z$.

Finally, we turn to a systematic exploration of hysteresis as a function of temperature sweeps at fixed $H_z$ [experimental sequence, Fig.~\ref{fig:4}(a)].
We begin by zero field cooling the sample and then ramping up a magnetic field to various strengths. 
We measure ODMR spectra as the temperature is increased above $T_\textrm{c}$ (field heating) and then decreased below $T_\textrm{c}$ (field cooling). 
The resulting data for $B_z$ are illustrated in Fig.~\ref{fig:5}(d).
Two features are apparent in the data. 
First, the hysteresis between field heating (red) and field cooling (blue) is enhanced at larger $H_z$ fields.
%
%
Second, at $H_z=206$~G, we observe a surprising sharpening of the magnetic transition (measured via the jump in $B_z$) compared to that seen at smaller $H_z$ fields.
%


\emph{Discussion and Conclusions}---Our demonstration of magnetic imaging up to $\sim140$~GPa with high contrast and minimal stress inhomogeneity opens up a new range of measurements on materials in the megabar regime, including both DC magnetometry and noise spectroscopy~\cite{xu2023recent,casola2018probing}.
Looking forward, our work opens the door to a number of intriguing directions. 
First, we expect that our NV-based quantum sensing techniques can readily be extended to even higher pressures (i.e.~$\gtrsim200$~GPa), e.g. using double-beveled anvils~\cite{o2018contributed}.
Second, it would be interesting to revisit other high pressure superconductors, such as LaH\textsubscript{10}, H\textsubscript{3}S 
and the recently unveiled N-Lu-H~\cite{dasenbrock2023evidence,li2023superconductivity, shan2023pressure}, where prior magnetic measurements have been limited to global probes, such as SQUIDs \cite{drozdov2015conventional, minkov2206trapped} and pick-up coils \cite{huang2019high, struzhkin2020superconductivity}. 
%
%
%
Third, local magnetometry is particularly important for superhydride materials, where synthesis by laser heating leads to inhomogeneities at the micron-scale. 
Imaging and characterizing such inhomogeneities is a crucial step toward quantifying sample yield and improving synthesis recipes.

While our results corroborate the qualitative picture of superconductivity argued for in related superhydride systems, extracting the properties of an idealized ``parent'' superconductor  is non-trivial. 
To this end, we expect widefield NV microscopy~\cite{scholten2021widefield} to allow for an efficient investigation of  the effects of the size and distribution of superconducting grains.
Coupled with noise spectroscopy, this would enable simultaneous measurements of the spatial and temporal current fluctuations in superconductors~\cite{dolgirev2022characterizing, chatterjee2022single}.

\vspace{3mm}

\emph{Acknowledgments}---This work was supported as part of the Center for Novel Pathways to Quantum Coherence in Materials, an Energy Frontier Research Center funded by the U.S. Department of Energy, Office of Science, Basic Energy Sciences under Award no. DE-AC02-05CH11231. 
C. R. L. acknowledges support from the National Science Foundation (grant PHY-1752727).
M.B. acknowledges support from the Department of Defense (DoD) through the National Defense Science and Engineering Graduate (NDSEG) Fellowship Program.
S.H. and S.M. acknowledges support from the National Science Foundation Graduate Research Fellowship under Grant No. DGE-1752814.
N.Y.Y. acknowledges support from the David and Lucile Packard foundation.

\vspace{3mm}


\vspace{3mm}


\emph{Data availability}---The data presented in this study are available from the corresponding author on request.

\bibliography{main.bib}

\newpage

\newpage

\newpage

\newpage

\newpage

\end{document}


\title{Supplementary information: Imaging magnetism in a hydride superconductor at megabar pressures using nanoscale quantum sensor}

\author{Prabudhya Bhattacharyya}
\thanks{These authors contributed equally to this work.}
\affiliation{Department of Physics, University of California, Berkeley, CA 94720, USA}
\affiliation{Materials Science Division, Lawrence Berkeley National Laboratory, Berkeley, CA 94720, USA}

\author{Wuhao Chen}
\thanks{These authors contributed equally to this work.}
\affiliation{State Key Laboratory of Superhard Materials, College of Physics, Jilin University, Changchun 130012, China}

\author{Xiaoli Huang}
\thanks{These authors contributed equally to this work.}
\affiliation{State Key Laboratory of Superhard Materials, College of Physics, Jilin University, Changchun 130012, China}

\author{Shubhayu Chatterjee}
\affiliation{Department of Physics, University of California, Berkeley, CA 94720, USA}
\affiliation{Department of Physics, Carnegie Mellon University, Pittsburgh, PA 15213, USA}

\author{Benchen Huang}
\affiliation{ Department of Chemistry, University of Chicago, Chicago, IL 60637, USA}

\author{Bryce Kobrin}
\affiliation{Department of Physics, University of California, Berkeley, CA 94720, USA}
\affiliation{Materials Science Division, Lawrence Berkeley National Laboratory, Berkeley, CA 94720, USA}

\author{Yuanqi Lyu}
\affiliation{Department of Physics, University of California, Berkeley, CA 94720, USA}

\author{Thomas J. Smart}
\affiliation{Department of Physics, University of California, Berkeley, CA 94720, USA}
\affiliation{Department of Earth and Planetary Science, University of California, Berkeley, CA 94720, USA}
\author{Maxwell Block}
\affiliation{Department of Physics, Harvard University, Cambridge, MA 02135, USA}

\author{Esther Wang}
\affiliation{Department of Chemistry and Chemical Biology, Harvard University, Cambridge, MA 02135, USA}

\author{Zhipan Wang}
\affiliation{Department of Physics, Harvard University, Cambridge, MA 02135, USA}

\author{Weijie Wu}
\affiliation{Department of Physics, Harvard University, Cambridge, MA 02135, USA}

\author{Satcher Hsieh}
\affiliation{Department of Physics, University of California, Berkeley, CA 94720, USA}
\affiliation{Materials Science Division, Lawrence Berkeley National Laboratory, Berkeley, CA 94720, USA}

\author{He Ma}
\affiliation{ Department of Chemistry, University of Chicago, Chicago, IL 60637, USA}

\author{Srinivas Mandyam}
\affiliation{Department of Physics, Harvard University, Cambridge, MA 02135, USA}

\author{Bijuan Chen}
\affiliation{Department of Physics, Harvard University, Cambridge, MA 02135, USA}

\author{Emily Davis}
\affiliation{Department of Physics, University of California, Berkeley, CA 94720, USA}

\author{Zachary M. Geballe}
\affiliation{Earth and Planets Laboratory, Carnegie Institution of Washington, Washington DC 20015, USA}

\author{Chong Zu}
\affiliation{Department of Physics, Washington University in St. Louis, St. Louis, MO 63130, USA}

\author{Viktor Struzhkin}
\affiliation{Center for High Pressure Science and Technology Advanced Research, Shanghai 201203, China}

\author{Raymond Jeanloz}
\affiliation{Department of Earth and Planetary Science, University of California, Berkeley, CA 94720, USA}

\author{Joel E. Moore}
\affiliation{Department of Physics, University of California, Berkeley, CA 94720, USA}
\affiliation{Materials Science Division, Lawrence Berkeley National Laboratory, Berkeley, CA 94720, USA}

\author{Tian Cui}
\affiliation{State Key Laboratory of Superhard Materials, College of Physics, Jilin University, Changchun 130012, China}
\affiliation{School of Physical Science and Technology, Ningbo University, Ningbo 315211, China}

\author{Giulia Galli}
\affiliation{ Department of Chemistry, University of Chicago, Chicago, IL 60637, USA}
\affiliation{Materials Science Division and Center for Molecular Engineering, Argonne National Laboratory, Lemont, IL 60439, USA}
\affiliation{Pritzker School of Molecular Engineering, University of Chicago, Chicago, IL 60637, USA}

\author{Bertrand I. Halperin}
\affiliation{Department of Physics, Harvard University, Cambridge, MA 02135, USA}

\author{Chris R. Laumann}
\affiliation{Department of Physics, Boston University, Boston, MA 02215, USA}

\author{Norman Yao}
\affiliation{Department of Physics, University of California, Berkeley, CA 94720, USA}
\affiliation{Materials Science Division, Lawrence Berkeley National Laboratory, Berkeley, CA 94720, USA}
\affiliation{Department of Physics, Harvard University, Cambridge, MA 02135, USA}

\date{\today}
\maketitle
\section{The NV center}
The NV center in diamond is a crystallographic defect comprising a substitutional nitrogen atom adjacent to a lattice vacancy~\cite{doherty2013nitrogen}. 
The electronic ground state of the negatively charged NV center is spin triplet ($S~=~1$). 
In the absence of external perturbations, the ground-state spin Hamiltonian is given by $H= D_{\textrm{gs}}S_z^2$, where $D_{\textrm{gs}}~=~(2\pi)~\times~2.87$~GHz is a temperature-dependent zero-field splitting between the $\ket{m_\textrm{s}=0}$ spin sub-level and the degenerate $\ket{m_\textrm{s}=\pm1}$ spin sub-levels, and $\{S_x, S_y, S_z\}$ are spin-1 operators quantized along the N-V axis~($\hat z$). 
The quantization axis may be oriented along any of the diamond bonds resulting in four subgroups of NV centers.
A magnetic field $\vec B$ couples with the Hamiltonian $H_B=\gamma_B\vec B\cdot\vec S$, where $\gamma_B=(2\pi)\times2.8$~MHz/G is the NV's gyromagnetic ratio and $\vec B$ is expressed in the local frame of the NV center~\cite{doherty2013nitrogen,maze2008nanoscale}. 

A stress tensor, $\pmb{\sigma}$, couples with the Hamiltonian~\cite{hsieh2019imaging,barson2017nanomechanical},
\begin{equation}
H_\textrm{s}=\Pi_z S_z^2+\Pi_x (S_y^2-S_x^2)+\Pi_y (S_xS_y+S_yS_x)
\end{equation}
where $\Pi_i = \Pi_i(\pmb{\sigma})$ are functions of the stress tensor as follows:
\begin{align}
    \Pi_z &= \alpha_1 \big(\sigma_{xx} + \sigma_{yy}\big) + \beta_1 \sigma_{zz}\\
    \Pi_x &= \alpha_2 \big(\sigma_{yy}- \sigma_{xx}\big) + \beta_2 \big(2 \sigma_{xz}\big)\label{eq:Ex}\\
    \Pi_y &= \alpha_2 \big(2 \sigma_{xy}\big) + \beta_2 \big(2\sigma_{yz}\big)\label{eq:Ey}
\end{align} 
are expressed in terms of the stress components in the NV's local frame.
Here, $\{\alpha_1, \beta_1, \alpha_2 , \beta_2 \}= (2\pi)\times \{8.6(2), -2.5(4), -1.95(9), -4.50(8)\} \; \SI{}{\mega \hertz/\giga \pascal}$ are the stress susceptibilities.
Stress components that preserve the C\textsubscript{3v}-symmetry of the NV center maintain the degeneracy of the $\ket{m_{\textrm{s}}=\pm1}$ spin-sublevels and induce an overall energy shift of $\Pi_z$. 
Stress components that break C\textsubscript{3v}-symmetry lift the degeneracy between the $\ket{m_{\textrm{s}}=\pm1}$ spin-sublevels and induce a splitting $2\Pi_\perp=2\sqrt{\Pi_x^2+\Pi_y^2}$.
In addition, symmetry breaking stress components mix the $\ket{m_\textrm{s}=\pm1}$ states into eigenstates $\ket{\pm}= (\ket{m_{\textrm{s}}=1} \pm e^{i\phi_{\Pi}}\ket{m_{\textrm{s}}=-1})/\sqrt{2}$, where $\phi_{\Pi}= \arctan(\Pi_y/\Pi_x)$. 

We excite electronic transitions in the NV center using laser light with wavelength $\lambda=532$~nm. 
Radiative transitions involved in electronic excitation and fluorescent relaxation in the NV center are spin conserving processes~\cite{doherty2013nitrogen, goldman2015state}. 
In addition to these radiative transitions, the NV center also shows a non-radiative inter system crossing (ISC) mechanism, whereby the $\ket{m_\textrm{s}=\pm1}$ spin sub-levels of the electronic excited state can transition into the singlet manifold. 
Further non-radiative relaxation from the singlet manifold to the electronic ground state is not highly spin selective~\cite{doherty2013nitrogen, goldman2015state}.
This leads to two important consequences.
First, it enables optical polarization to the $\ket{m_\textrm{s}=0}$ spin sub-level under continuous laser excitation. 
Second, it causes a reduction in the fluorescence rates of the $\ket{m_\textrm{s}=\pm1}$ spin sub-levels with respect to the $\ket{m_\textrm{s}=0}$ spin sub-level. 
Indeed, signal contrast of an ODMR measurement depends on the difference between the fluorescence intensity of these spin sub-levels.
As we will discuss next, the ISC mechanism --- and thus the ODMR contrast --- is sensitive to the applied stress and can be significantly altered at high pressures.

\subsection{NV sensing at megabar pressures}

\begin{figure}[h!]
         \centering
         \includegraphics[width = \textwidth]{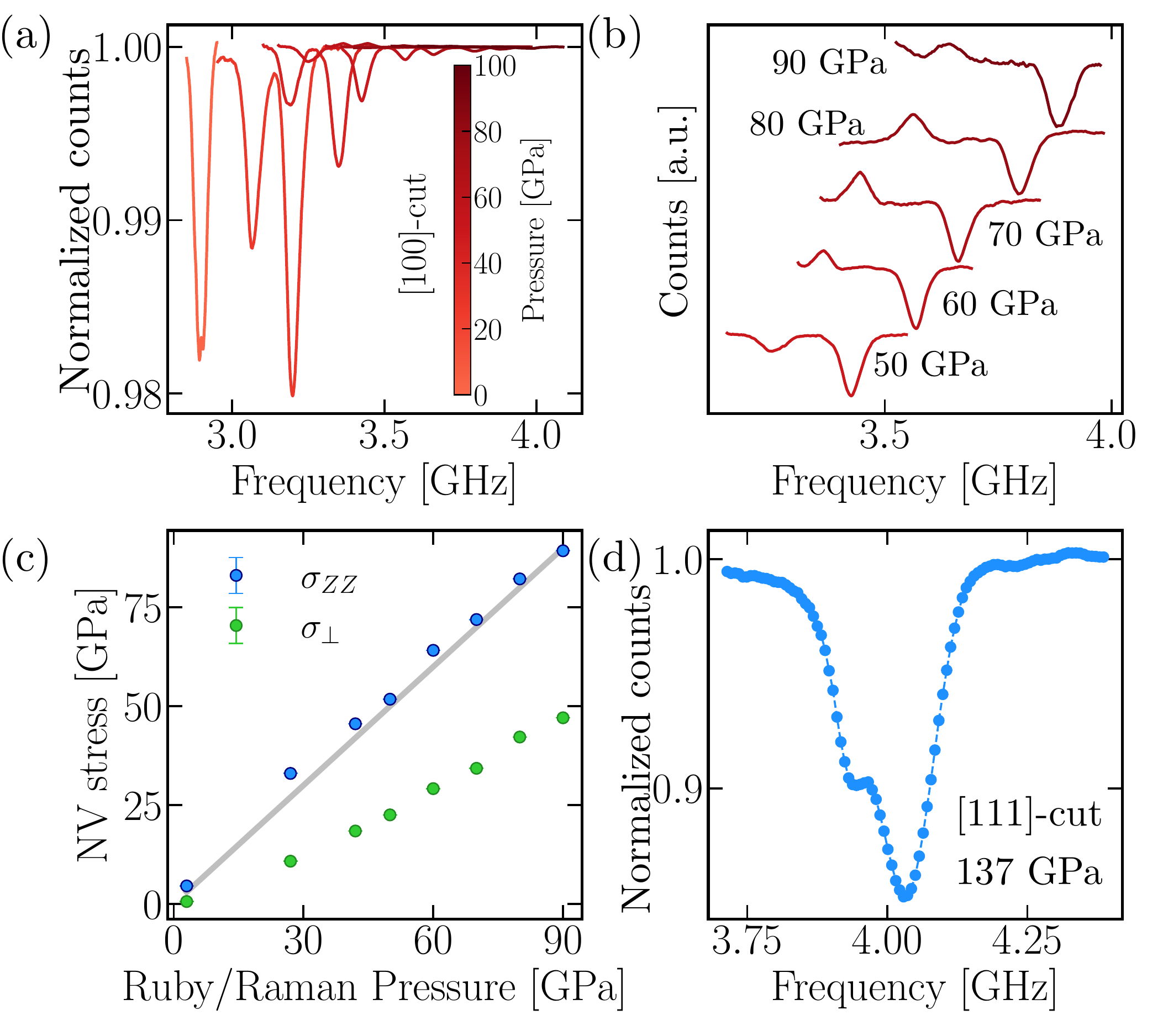}
         \caption{
           \textbf{Pushing NV sensing to megabar pressures.}
           \textbf{(a)} ODMR measurements under pressure in a [100]-cut anvil exhibit a drastic reduction in contrast with increase in pressure.
           The dominant culet stresses $\{\sigma_{ZZ}, \sigma_{\perp}\}$ have degenerate symmetry-preserving and symmetry-breaking projections on each NV subgroup, thereby inducing both a shift, $\Pi_z$, and a splitting, $2\Pi_{\perp}$, with increasing pressure. 
           \textbf{(b)} In a [100]-cut anvil, we see a surprising inversion of contrast for one of the ODMR peaks at pressures above $\sim50$~GPa. 
           \textbf{(c)} In a [100]-cut anvil, we see good agreement between the values of the loading stress $\sigma_{ZZ}$ extracted from NV measurements (blue points) and values of the sample pressure calibrated via a combination of ruby and culet raman spectra~\cite{syassen2008ruby,akahama2006pressure}.
           For comparison, $x=y$ line is plotted in grey.
           We also measure an increase in the $\sigma_{\perp}$ stress (green points) consistent to cupping of the diamond culet~\cite{li2018diamond}. 
           \textbf{(d)} In a [111]-cut culet (sample S2), we show $\sim15\%$ contrast at $137$~GPa pressure for the [111] NV subgroup.
           }
         \label{fig:100-study}
\end{figure}

The extension of NV sensing to megabar pressures has been a long-standing challenge due to several factors: reduction in NV fluorescence, inhomogenous broadening of ODMR spectra, and significant loss of ODMR contrast ~\cite{doherty2014electronic,dai2022optically}. 
In this section, we compare NV measurements under pressure for [100]-cut and [111]-cut anvils [Fig.~\ref{fig:100-study}].  
%

For studies in a [100]-cut anvil, we use a type Ib diamond with \SI{200}{\micro\meter} flat culet having a shallow layer of NV centers incorporated up to $50$~nm below the surface (recipe included in the following section)~\cite{hsieh2019imaging}. 
We use an untreated type Ia diamond as the opposing anvil, a potassium chloride (KCl) pressure medium, and a combination of ruby and culet raman spectra for pressure calibration~\cite{syassen2008ruby, akahama2006pressure}.
We refer to the lab frame by $\{X,Y,Z\}$ where $Z$ is the loading axis, and to the local frame of the NV subgroups by $\{x,y,z\}$.

Using the NV center, we measure the dominant culet stresses, $\sigma_{ZZ}$ and $\sigma_{\perp}$. 
The former, $\sigma_{ZZ}$, is the loading stress, and the latter, $\sigma_{\perp}=(\sigma_{XX} + \sigma_{YY})/2$, is a biaxial in-plane stress, which at high pressures is related to cupping of the diamond culet~\cite{li2018diamond}. 
For a [100]-cut anvil, these stresses project degenerately across all four NV subgroups, having both symmetry preserving and symmetry breaking components (Fig.~1(b) in maintext).
In particular, the measured stress parameters $\{\Pi_z,\Pi_\perp \}$ map to $\{\sigma_{ZZ}, \sigma_{\perp}\}$ stresses as follows: 
\begin{align}
    \Pi_z &= \frac{1}{3}(2\alpha_1+\beta_1)(2\sigma_{\perp} + \sigma_{ZZ})\\
    \Pi_\perp &= \frac{2}{3}|(\alpha_2 - {\sqrt{2}}\beta_2) (\sigma_{ZZ}-\sigma_{\perp})|
\end{align}
We use this relationship to extract the stresses from the NV's ODMR spectrum measured at the center of the culet [Fig.~\ref{fig:100-study}(c)]. 
The loading stress, $\sigma_{ZZ}$, agrees closely with the sample pressure which is calibrated independently (using ruby and culet raman pressure markers)~\cite{syassen2008ruby, akahama2006pressure}. 

\emph{ODMR contrast}---For the [100]-cut anvil, we observe a significant loss in the ODMR contrast with increasing pressure~[Fig.~\ref{fig:100-study}(a)].
In addition, at pressures above~$\sim50$~GPa, we observe a surprising inversion of ODMR contrast for one of the resonances~[Fig.~\ref{fig:100-study}(b)]. 
However, for the [111]-cut anvil, neither of these effects are observed.
Notably, for this anvil cut, the ODMR contrast at pressures up to $\sim140$~GPa remains comparable to that typically achieved under ambient conditions~[Fig.~\ref{fig:100-study}(d)].

A microscopic understanding of the observed changes in the ODMR contrast requires a careful analysis of the NV center's ISC mechanism in the presence of stress~\cite{goldman2015state}.
While this is the subject of ongoing studies~\cite{bencheninprep}, we summarize our preliminary findings as follows.
C\textsubscript{3v}-symmetry preserving stresses induce an overall quantitative change in the ISC rate.
Interestingly, a mixture of hydrostatic stress and an uniaxial stress along the quantization axis ($\hat z$) of the NV center leads to the highest ISC rate at megabar pressures.
This condition is approximated in a [111]-cut anvil for one of the NV subgroups and may explain the retention of ODMR contrast at these pressures [Fig.~\ref{fig:100-study}(d)]~\cite{hsieh2019imaging}.
In comparison, C\textsubscript{3v}-symmetry breaking stresses can allow new non-radiative ISC transitions from the spin manifold in the electronic excited state, thereby inducing qualitatively different changes to the ISC mechanism.
In a [100]-cut anvil, the presence of symmetry breaking stresses may be responsible for both contrast loss as well as the appearance of positive contrast peaks.

\section{Experimental details}
\subsection{Sample preparation}
For NV measurements, we use 16-sided standard design diamond anvils with [111]-crystal cut culets (Syntek Co. Ltd. and Almax-easyLab). 
These anvils are polished from synthetic type-Ib ([N]~$\leq200$~ppm) single crystal diamonds. 
The culet diameter is \SI{100}{\micro\meter} with a bevel at \SI{8.5}{\degree} up to \SI{300}{\micro\meter}.
Similar to previous work, we perform \textsuperscript{12}C\textsuperscript{+} ion implantion at an energy of $30$~keV with a dosage of $5\times10^{12}$~cm\textsuperscript{-2} (CuttingEdge Ions, LLC) to generate a layer of vacancies up to $50$~nm from the culet surface~\cite{hsieh2019imaging}.
Following implantation, we vacuum anneal the diamond anvils (at $P<10^{-6}$~mbar) in a home built furnace at a temperature $>$~\SI{850}{\celsius} for $12$~hours. 
During the annealing step, the vacancies become mobile in the lattice and can probabilistically form NV centers by migrating to lattice sites adjacent to existing substitutional nitrogen defects. 
The NV diamond is glued to the cylinder side of a $23$-mm miniature panoramic diamond anvil cell (DAC) machined out of non-magnetic NiCrAl alloy and beryllium copper (Cu-Be)~\cite{gavriliuk2009miniature}. 
A platinum (Pt) wire with a diameter of  \SI{20}{\micro\meter} is incorporated near the culet to apply microwave (MW) radiation for spin state control. During compression, the Pt wire on the diamond bevel gradually expands into a foil.
On the opposing (piston) side, we glue a type-Ia diamond anvil with [100]-crystal cut.
On this anvil, we sputter Molybdenum (Mo) leads for performing electrical resistance measurements. 
A double-sided insulating gasket, fashioned out of rhenium (Re) foil and calcium fluoride (CaF\textsubscript{2}) mixed with epoxy, is used to isolate the transport leads and the MW wire.

\begin{figure}[h!]
         \centering
         \includegraphics[width = \textwidth]{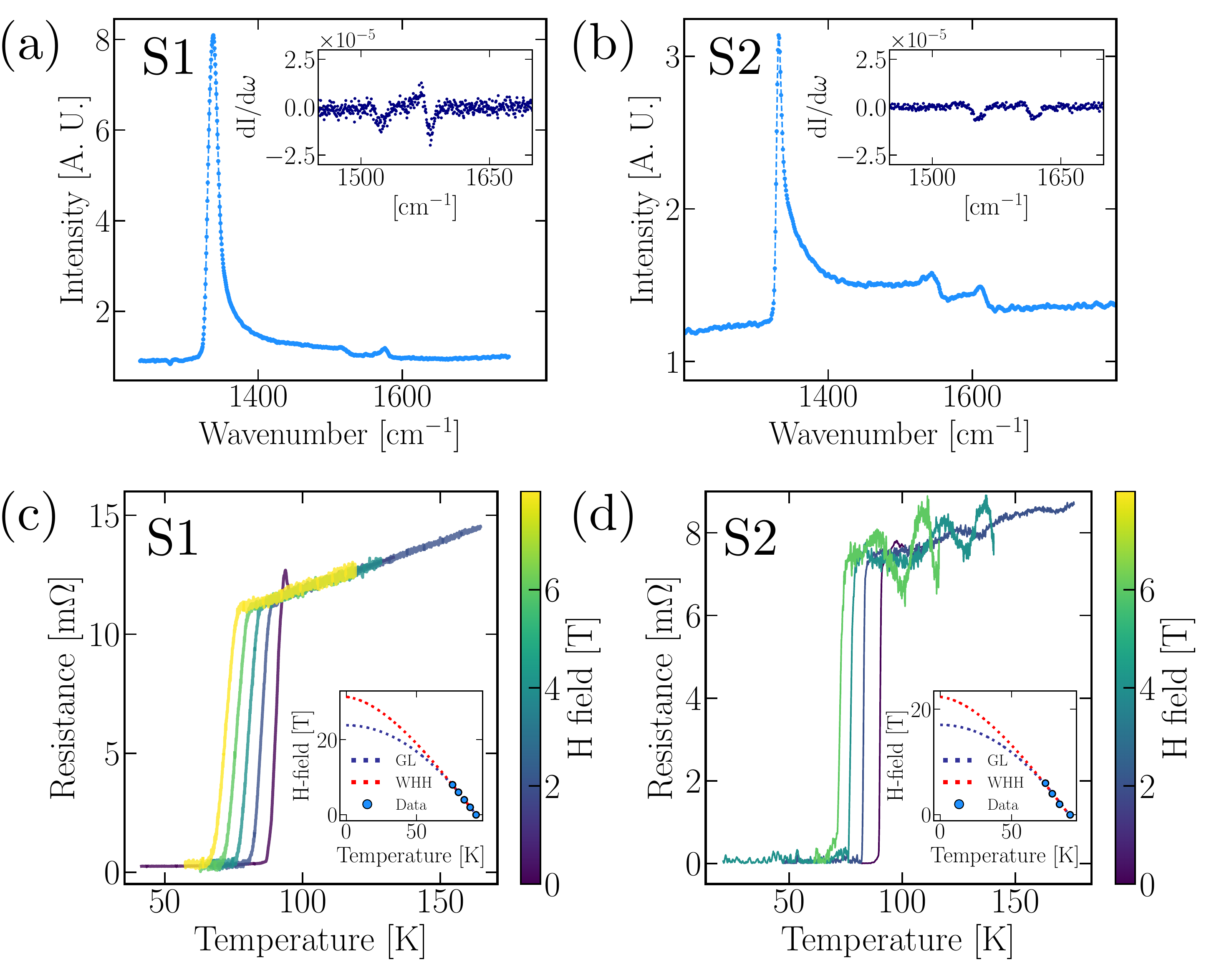}
         \caption{
           \textbf{Pressure calibration and verification of CeH\textsubscript{9} synthesis. }
           Following the procedure detailed in \cite{akahama2005raman}, we use culet raman spectra for samples \textbf{(a)} S1 and \textbf{(b)} S2 to determine loading pressure. 
           The insets plot the derivative of the raman fluorescence intensity; we ascertain pressures of $\sim 118$~GPa for sample S1 and $\sim137$~GPa for sample S2 respectively. 
           Following synthesis through laser heating, four point resistance of \textbf{(c)} S1 and \textbf{(d)} S2 were measured via delta technique~\cite{handbook20237th}.
           The measured $T_\textrm c=94$~K for S1 and $T_\textrm c=91$~K for S2
           suggest the formation of CeH\textsubscript{9} in both samples~\cite{chen2021high}. 
           $T_\textrm c$ is determined by onset of the resistance drop.
           Insets depict the dependence of $T_\textrm c$ on the applied field $H$ (blue data points). 
           Within the parameter regime of operation, the measurements show good fits to both Ginzburg-Landau (GL) theory (blue dotted line)~\cite{ginzburg1950j} and Werthamer-Helfand-Hohenberg (WHH) theory (red dotted line)~\cite{werthamer1966temperature,baumgartner2013effects}.
           }
         \label{fig:raman-and-transport}
\end{figure}

We synthesize cerium hydride following the recipe detailed in~\cite{chen2021high}. 
We load cerium (Ce) metal (99.8\% purity) and ammonia borane (NH\textsubscript{3}BH\textsubscript{3}) in an argon (Ar) glove box.
We first compress the loading to pressures above $100$~GPa.
The pressure $P$ is calibrated using the raman line of the [111]-cut culet [see Fig.~\ref{fig:raman-and-transport}(a,b)]~\cite{akahama2005raman}.
Following this, we perform single-sided laser heating by focusing a $1070$~nm laser beam to a $\sim3$~\SI{}{\micro\meter} spot at different locations on the Ce sample. 
Laser heating induces the decomposition of ammonia borane into cubic boron nitride (cBN) and hydrogen. 
The released hydrogen subsequently reacts with the cerium to form cerium polyhydrides. 
(Ce + NH\textsubscript{3}BH\textsubscript{3}~$\rightarrow$~\textrm{CeH}$_x$~+ cBN).
After synthesis, four point resistance is measured using the delta technique~\cite{handbook20237th} in a liquid He (wet) cryostat (detailed in~\cite{chen2021high}).
$T_\textrm{c}$ is determined by the onset of the resistance drop.
We measure $T_\textrm{c}\approx92$~K for S1 at $P\approx118$~GPa and $T_\textrm{c}\approx91$~K for S2 at $P\approx137$~GPa indicating the formation of CeH\textsubscript{9} in both samples [see Fig.~\ref{fig:raman-and-transport}(c,d)].
$T_\textrm{c}$ is suppressed upon the application of tesla scale magnetic fields [insets in Fig.~\ref{fig:raman-and-transport}(c,d)].
For both samples S1 and S2, we find that the change in $T_\textrm{c}$ is less than $0.5\%$ for an applied field $H<1000$~G. 
This allows us to calibrate for temperature gradients in the dry cryostat where NV measurements are performed up to $H_z=240$~G fields [Fig.~\ref{fig:T-gradient}].
We note that sample S1 failed before all studies could be completed.

\subsection{Apparatus for simultaneous ODMR and electrical resistance measurements}
 
We perform NV measurements in a dry closed cycle cryostat (Attocube AttoDRY800) integrated with home-built confocal fluorescence microscope (similar to previous work~\cite{hsieh2019imaging}). 
A galvanometer (Thorlabs GVS212) is used to raster the excitation beam from a $532$~nm laser (Sprout Solo-10W). 
The excitation light is switched by an acousto-optic modulator (AOM, Gooch \& Housego AOMO 3110-120) in double pass configuration.
A $4f$-telescope focuses the laser beam at the entrance of an objective lens (Mitutoyo LWD NA 0.7) that is used to image the diamond culet.
The fluorescence collected by the objective is separated from the excitation path by a dichroic mirror and measured by a fiber-coupled single photon counting module (Excelitas SPCM-AQRH-64FC).
A data acquisition device (National Instruments USB-6343) is used for fluorescence counting as well as modulation of a microwave source (SRS SG386) for continuous wave ODMR measurements.
For pulsed measurments, a programmable multi-channel pulse generator (SpinCore PulseBlasterESR-PRO 500) is used to gate the AOM switch and the microwave source (SRS SG386).
For both continous wave and pulsed measurments, the microwaves are amplified using a Minicircuits ZHL-25W-63+ amplifier and subsequently channeled to the sample using coaxial cables.
The Pt-wire at the sample is connected to the coaxial terminals using \SI{100}{\micro\meter} copper magnet wires.
An external magnetic field is applied by running a current $I$ from a DC power supply (BK Precision 9103) into a custom built electromagnet (Woodruff Scientific) that sits outside the cryostat shroud [Fig.~\ref{fig:Hz-calibration}(a)].
A lock-in amplifier (Zurich Instruments MLFI DC-500kHz) in combination with a voltage-controlled current source (Stanford Research Systems CS580) is used to measure four point resistance of the sample.
In this setup, we are limited to applied fields of $\approx240$~G. 
Electrical resistance measurements at high fields [Fig.~\ref{fig:raman-and-transport}(c,d)] are performed in a separate cryostat~\cite{chen2021high}.

In sample S2, we find that CeH\textsubscript{9} region is synthesized directly on top of two transport leads [Fig.~\ref{fig:laser-heating-NV-creation}(c)]. 
For four point resistance measurements, we pick up voltage between these two transport leads to get near-zero resistance for $T<T_\textrm{c}$.

\subsection{Creation of NV centers during laser heating}

\begin{figure}[h!]
         \centering
         \includegraphics[width = \textwidth]{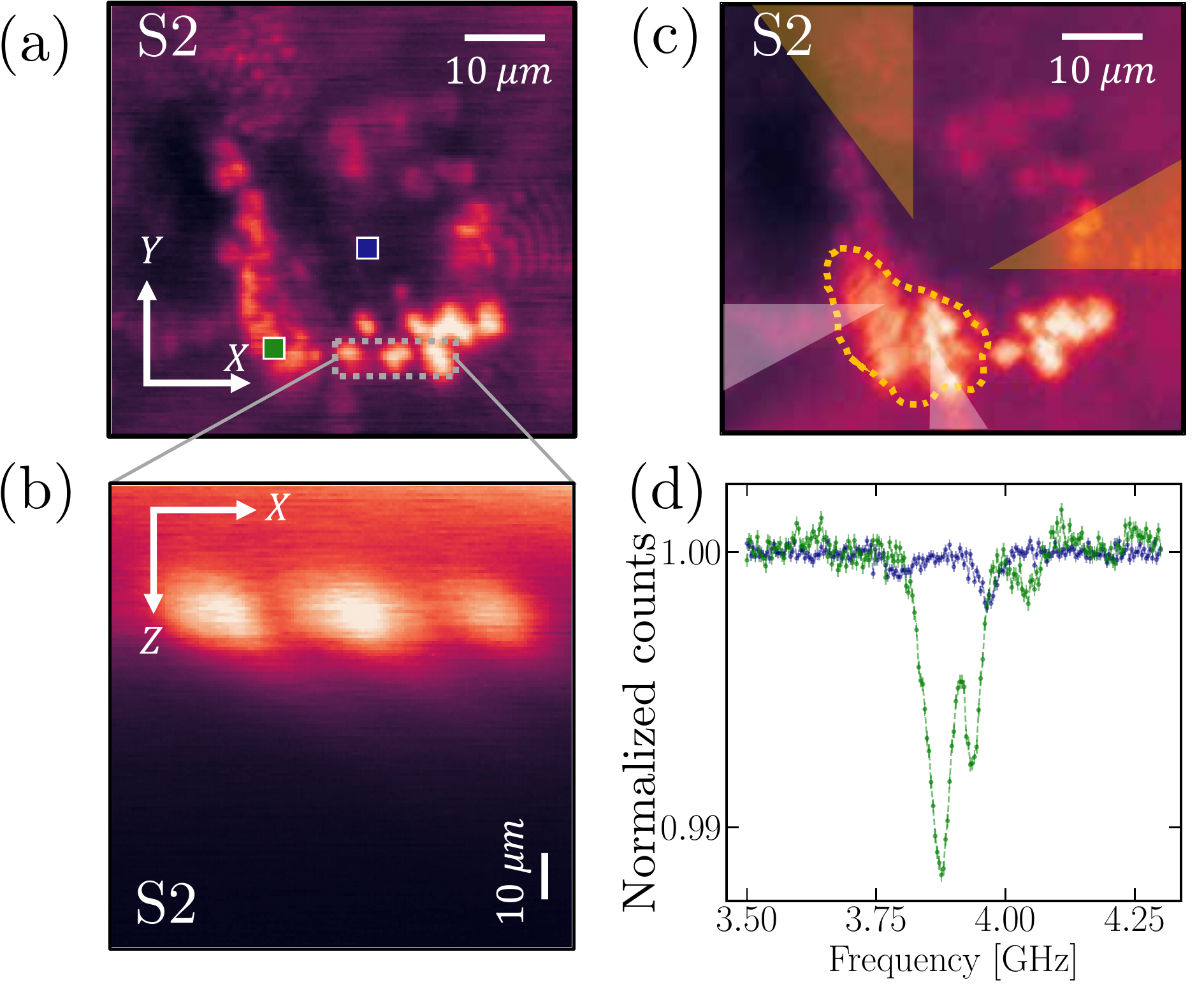}
         \caption{
           \textbf{Creation of NV centers at the culet during to laser heating.}
           \textbf{(a)} A confocal image of sample S2 after an initial round of laser heating shows bright regions with higher NV fluorescence suggesting the creation of NV centers during laser heating for superhydride synthesis~\cite{chen2017laser}.
           %
           %
           \textbf{(b)} A confocal image in the $XZ$ plane of the diamond culet suggests that NV centers created during laser heating extend deeper into the anvil ($\sim$ microns) compared to NV centers created by implantation and annealing ($\sim50$~nm).
           \textbf{(c)} A second round of laser heating of sample S2 in the region enclosed by a dotted yellow line clearly shows the creation of more NV centers in this region [compared to scan in (a)].
           This region sits on top of two of the transport leads (white triangles). 
           In four point resistance measurements we pick up voltage in these two leads and run current through the other two (yellow triangles). 
           \textbf{(d)} A comparison of an ODMR spectrum (green) measured in the laser heated region [green point in (a)] to the spectrum (blue) measured outside the laser heated region [blue point in (a)]. 
           }
         \label{fig:laser-heating-NV-creation}
\end{figure}

In addition to the layer of NV ensembles created by ion implantation and annealing, we find that a dense ensemble of NV centers are created by laser annealing of the diamond during CeH\textsubscript{9} synthesis.
In particular, regions of the sample that are laser heated show noticeably higher fluorescence in both samples S1 and S2.
Fig.~\ref{fig:laser-heating-NV-creation} illustrates this for sample S2: we see micron-scale bright spots resembling foci of an optical beam. 
A confocal scan along the vertical depth ($Z$) into the diamond [Fig.~\ref{fig:laser-heating-NV-creation}(c)] suggests that the NV centers created due to laser annealing are distributed up to a few microns into the anvil.

We note the following remarks. 
First, due to compression of the sample to megabar pressures, we expect significant contact stress at the interface between the diamond and the Ce sample.
Indeed, we find that ODMR spectra in regions without laser heating exhibit large splittings ($2\Pi_{\perp}\gtrsim(2\pi)\times200$~MHz) [blue data in Fig.~\ref{fig:laser-heating-NV-creation}(d)]. 
However, ODMR spectra at laser heated regions show appreciably smaller splittings ($2\Pi_{\perp}\sim(2\pi)\times100$~MHz) [green data in Fig.~\ref{fig:laser-heating-NV-creation}(d)].
On the one hand, this is consistent with local stress relaxation due to high temperatures generated by laser heating~\cite{uts2013effect}.
On the other hand, the ODMR spectra in laser heated regions sample a larger distribution of the stress gradient (in $Z$) at the culet; this may be an additional cause for lower values of the overall $2\Pi_{\perp}$ splittings.
Second, laser heated regions of the culet show markedly higher ODMR contrast compared to other regions [Fig.~\ref{fig:laser-heating-NV-creation}(d)].

\subsection{Calibration of external magnetic field $H_z$}

\begin{figure}[h!]
         \centering
         \includegraphics[width = \textwidth]{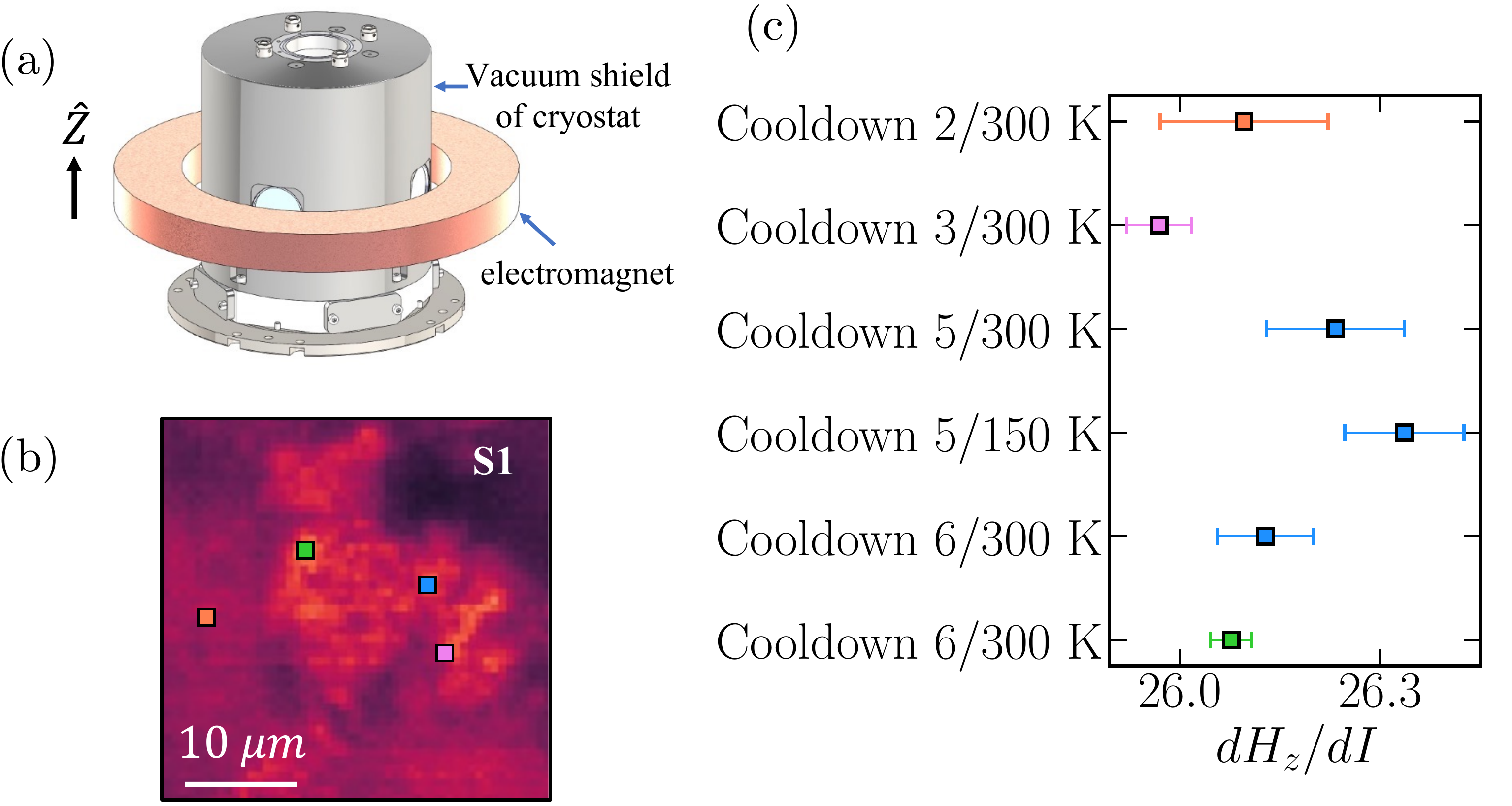}
         \caption{
           \textbf{Calibration of external magnetic field.}
           \textbf{(a)} The apparatus used for generating magnetic fields comprises a custom built electromagnet that sits outside the cryostat's vacuum shield.
           The electromagnet is oriented along loading axis, $\hat Z$, which is coincident with the quantization axis of the [111] NV subgroup, $\hat z$. 
           Due to minor changes in the placement of the electromagnet during experimental setup, we calibrate the magnetic field at the beginning of every cooldown. 
           \textbf{(b-c)} Calibrations at four spatial locations in sample S1 across four different cooldowns provide similar values for $dH_z/dI$. 
           For cooldown 5, calibrations in the superconducting region of the sample at two temperatures, $T=300$~K and $T=150$~K, yield similar results. 
           }
         \label{fig:Hz-calibration}
\end{figure}

At the beginning of every cooldown, we calibrate the external magnetic field ($H_z$) as a function of the current applied to the electromagnet ($I$).
We achieve this by measuring the NV's ODMR splitting $\Delta$ as a function of $I$. 
Our calibration necessarily assumes that, at temperatures $T>T_\textrm{c}$, the sample does not contribute to the local magnetic field, $B_z$, and thus $B_z= H_z$ (in Gaussian units).
Knowing $B_z=H_z\propto~I$ and $\Delta=\sqrt{(2\Pi_{\perp})^2 + (2\gamma_BB_z)^2}$, each calibration yields the proportionality constant $dH_z/dI$.
Due to the large size of the electromagnet coil ($\sim0.5$~m) compared to the diamond culet (\SI{100}{\micro\meter}), we expect the variation of applied field over our sample region to be small. [Fig.~\ref{fig:Hz-calibration}(a)].  
We verify this by performing the calibration at multiple spatial locations and also at several temperatures for the same spatial location over the course of several cooldowns [Fig.~\ref{fig:Hz-calibration}(b-c)]. 
Minor variations in the calibration ($dH_z/dI$) between cooldowns result from differences in the placement of the coil around the cryostat's vacuum shield during experimental setup.
Nevertheless, for each calibration we find that spatial variations in $H_z$ across the sample region is around $\sim1\%$ [see data for cooldown 6 in Fig.~\ref{fig:Hz-calibration}(c)]. 

\subsection{Correction for thermal gradients}

\begin{figure}[h!]
         \centering
         \includegraphics[width = \textwidth]{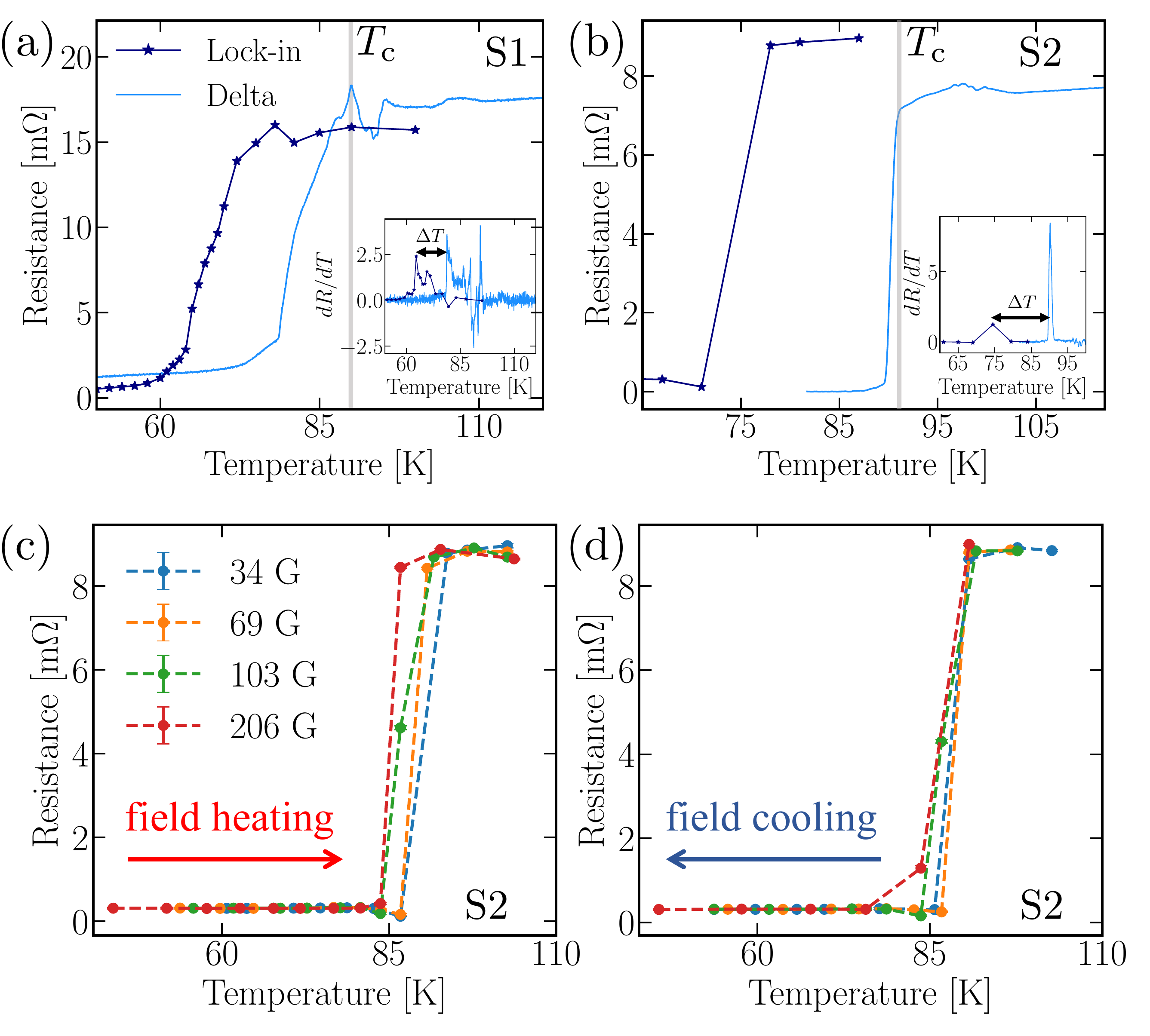}
         \caption{
           \textbf{Determination of thermal gradients.}
            Following synthesis, four point resistance of each samples is measured via delta technique [Fig.~\ref{fig:raman-and-transport}(c,d)] in a cryostat with good sample thermalization~\cite{chen2021high,handbook20237th}. 
            These measurements (shown in light blue data) for samples \textbf{(a)} S1 and \textbf{(b)} S2 are reflective of the true $T_\textrm c$ in each case.
            NV measurements are performed in a dry cryostat where sample thermalization can be challenging; in this cryostat, four point resistance of the sample is measured using lock-in technique (dark blue data).
            For both samples, the temperature at which we observe a resistive transition in the dry cryostat is lower than the true $T_\textrm c$, suggesting the presence of a temperature gradient ($\Delta T$).
            We extract $\Delta T$ by comparing the derivative ($dR/dT$) of sample resistance (measured via lock-in and delta techniques) [insets]. 
            Four point resistance of S2 measured via lock-in technique at several values of $H_z$ upon \textbf{(c)} field heating and \textbf{(d)} field cooling.
            The apparent spread in the transition points observed on field heating suggest that the temperature gradient in the dry cryostat may depend on the direction of the temperature sweep.
            A correction of $\Delta T=15.7$~K is applied to all data points in (c-d).
           }
         \label{fig:T-gradient}
\end{figure}

Following CeH\textsubscript{9} synthesis, four point resistance measured via delta technique in a Liq. He (wet) cryostat reflects the true $T_\textrm c$ for both samples [Fig.~\ref{fig:raman-and-transport}(c,d)].
NV measurements are performed in a dry closed cycle cryostat where the temperature is measured at the cold finger and not directly on the DAC body.
In dry cryostats, sample thermalization can pose a significant challenge due to limited thermal contact and large thermal mass of the DAC. 
We incorporate two additional thermal links: the first, between the cell body and cold finger (custom fabricated by FourNineDesign) and, the second, between the gasket and the cell body (cut out of indium foil).
Despite these attempts, we measure a temperature gradient, $\Delta T$, between the temperature recorded at the cold finger and the true sample temperature.
For both samples S1 and S2, four point resistance measured via lock-in technique in the dry cryostat shows transitions at temperatures lower than the true $T_\textrm c$ [Fig.~\ref{fig:T-gradient}(a,b)].

We determine $\Delta T$ by measuring the shift of the resistance drop between the four point lock-in and delta measurements. 
This corresponds to a peak in $dR/dT$ as shown in the insets of Fig.~\ref{fig:T-gradient}(a,b).
We estimate $\Delta T\approx14.2$~K for S1 and $\Delta T$ between $\approx15.7$~K and $\approx21.2$~K for S2 across different cooldowns. 
For clarity, we present sample temperature $T$ in our data by simply adding this correction to the temperature read out at the cold finger.

We note the following remarks concerning the accuracy of this correction.
First, minor adjustments of the sample and the thermal link between cooldowns can change this temperature gradient (as seen in the case of S2).
In the case of S1, due to coupling between the transport leads and the Pt wire, four point resistance could only be measured when the sample was disconnected from the microwave circuitry used for ODMR measurements.
This isolation of the system may lower the temperature gradient, leading to different temperature corrections ($\Delta T$) for ODMR and electrical resistance measurements.
In Fig.~\ref{fig:S1-suppression-studes}(e), the apparent separation between the magnetic and resistive transitions is likely due to this reason.
Second, the temperature gradient may itself be temperature dependent and our correction does not account for this. 
Third, when sweeping temperatures across $T_\textrm c$, the rate of sample thermalization may depend on the direction of the sweep (i.e., field heating as opposed to field cooling) [see Fig.~\ref{fig:T-gradient}(c,d)]. 
Finally, during NV measurments in the dry cryostat, four point resistance is often measured under an applied field $H_z<300$~G. 
We neglect the modest suppression (less than $0.1\%$) of the true $T_\textrm c$ of the samples when estimating $\Delta T$ [see insets of Fig.~\ref{fig:raman-and-transport}(c,d)]. 
In future work, the incorporation of a layer of NV centers near the table side of the diamond anvil at the optical window of the DAC may allow for \emph{in situ} temperature calibration. 
Nevertheless, the presence of a temperature gradient does not affect the central observations in our work.

\subsection{The effect of temperature}

The temperature dependence of the zero-field splitting leads of a difference of $\approx(2\pi)\times7.3$~MHz in the value of $D_\textrm{gs}$ between room temperature ($T=300$~K) and cryogenic temperatures ($T=4$~K)~\cite{chen2011temperature}.
In comparison, we observe significant stress induced shifts ($\Pi_z>(2\pi)\times1$~GHz) and splittings ($2\Pi_{\perp}\sim(2\pi)\times100$~MHz) in the ODMR spectrum at megabar pressures.
Furthermore, given that we only interpret the ODMR splitting caused by stress and magnetic fields, we can exclude the direct effect of temperature in our measurements. 
However, this does not preclude indirect effects, including a temperature dependent change in the stress due to thermal expansion or contraction in the DAC body (discussed below). 
%
%
%

\section{Electrical resistance and magnetism in CeH\textsubscript{9}}

\subsection{Local diamagnetism in S1}

In the maintext, we show the ODMR spectra and extracted $B_z$ values for our study of the Meissner effect on zero field cooling of sample S2 [Fig.~3(a,b,e)].
We also show the extracted $B_z$ values for sample S1 [Fig.~3(f)].
Here, we include the ODMR spectra [Fig.~\ref{fig:S1-suppression-studes}(b,c)] that were used in the extraction of $B_z$ at the relevant spatial locations in S1 [green and purple points in Fig.~\ref{fig:S1-suppression-studes}(a)].

\subsection{Spatial studies of S1 and S2}

In the maintext, we show a spatially varying local suppression of the external field $H_z$ [Fig.~3~(g)] measured along a line cut in sample S1. 
Here, we include plots of $B_z$ as a function of $H_z$ at several spatial points along this line cut [Fig.~\ref{fig:S1-spatial-studies}(b)].
We present the same study for two orthogonal line cuts on the CeH\textsubscript{9} region of sample S2 [Fig.~\ref{fig:S2-spatial-studies}].

For spatial studies of local diamagnetism in S1 [Fig.~\ref{fig:S1-spatial-studies}], we recorded minor spatial drifts in our measurement apparatus. 
As a result, a number of ODMR spectra measured for $H_z=0$ could not be reliably fit to extract the stress splitting, $2\Pi_{\perp}$.
To ensure consistent analysis for this data set, we determine $2\Pi_{\perp}$ by a complementary technique. 
Specifically, at each spatial point, we first fit the measured splitting, $\Delta$, to the function $\Delta= \sqrt{(2\Pi_{\perp})^2 + (2 c I)^2}$, where $I$ is the current applied to the electromagnet and $(\Pi_{\perp},c)$ are fitting parameters.
We use this fitted $2\Pi_{\perp}$ value to extract the local field, $B_z$, as a function of the applied field $H_z$ in S1.
A comparison of the values of $B_z$ extracted using this technique to the values extracted using our prescription (given in the maintext) for similar studies in sample S2 does not show any significant differences.

\subsection{Temperature sweep in S1}

\begin{figure}[h!]
         \centering
         \includegraphics[width = \textwidth]{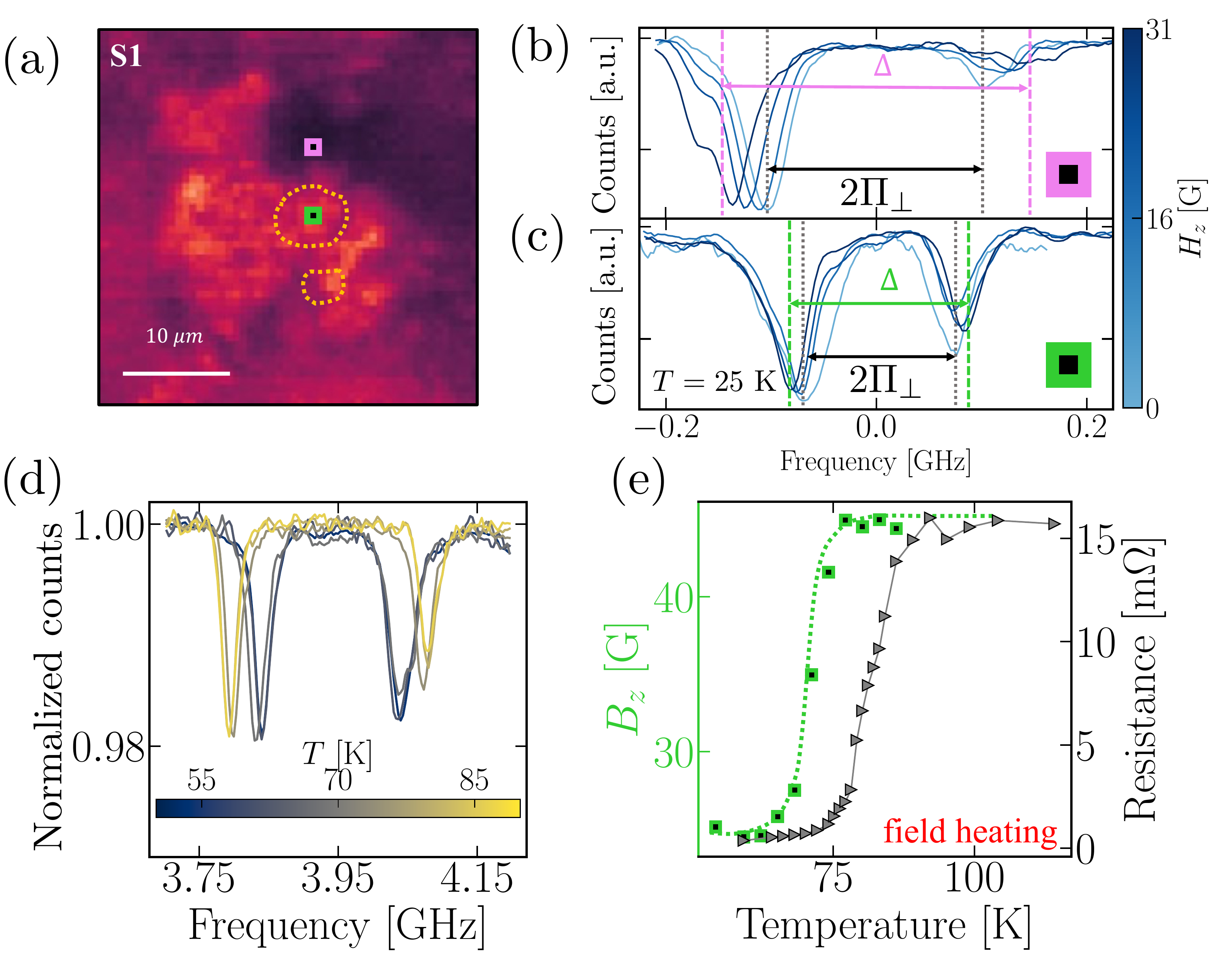}
         \caption{
           \textbf{Studies of diamagnetism in sample S1.}
           \textbf{(a)} Confocal fluorescence image of sample S1 showing the identified CeH\textsubscript{9} region (enclosed in dotted yellow line). 
           A comparison of the ODMR spectra measured after zero field cooling to a temperature $T=25$~K (below $T_\textrm c$) at two spatial locations: one \textbf{(b)} away from the CeH\textsubscript{9} region [purple point in (a)], and the other \textbf{(c)} on top of the CeH\textsubscript{9} region [green point in (a)].
           For clarity, all spectra are centered by subtracting the ODMR shift.
           As a function of the external field, $H_z$, the local field, $B_z$, extracted from these spectra show the diamagnetic response of CeH\textsubscript{9} [maintext Fig.~3(f)].
           We measure the temperature dependence of this response [at the green point in (a)]. (should the $\Pi_{\perp}$ in figure (b,c) be changed into $\Pi_{perp}$?)
           \textbf{(d)} Applying $H_z=47$~G after zero field cooling, we perform ODMR spectroscopy while heating the sample across $T_\textrm c$ (field heating).
           \textbf{(e)} Similar to sample S2 (shown in maintext), we observe a clear transition in the diamagnetic response.
           Four point resistance is measured separately on warming up the sample with $H_z=0$ (after zero field cooling).
           For sample S1, we are not able to perform simultaneous magnetometry and electrical resistance measurements due to coupling between the Pt wire for MW delivery and the transport leads. 
           Although, we expect a suppression in $T_\textrm c$ measured via electrical resistance on the application of an external field, the separation between the magnetic and resistive transitions is larger than expected from this effect alone.
           A reduction of the temperature gradient in the sample on disconnecting the MW lines may be responsible for this difference.
           }
         \label{fig:S1-suppression-studes}
\end{figure}

\begin{figure}[h!]
         \centering
         \includegraphics[width = \textwidth]{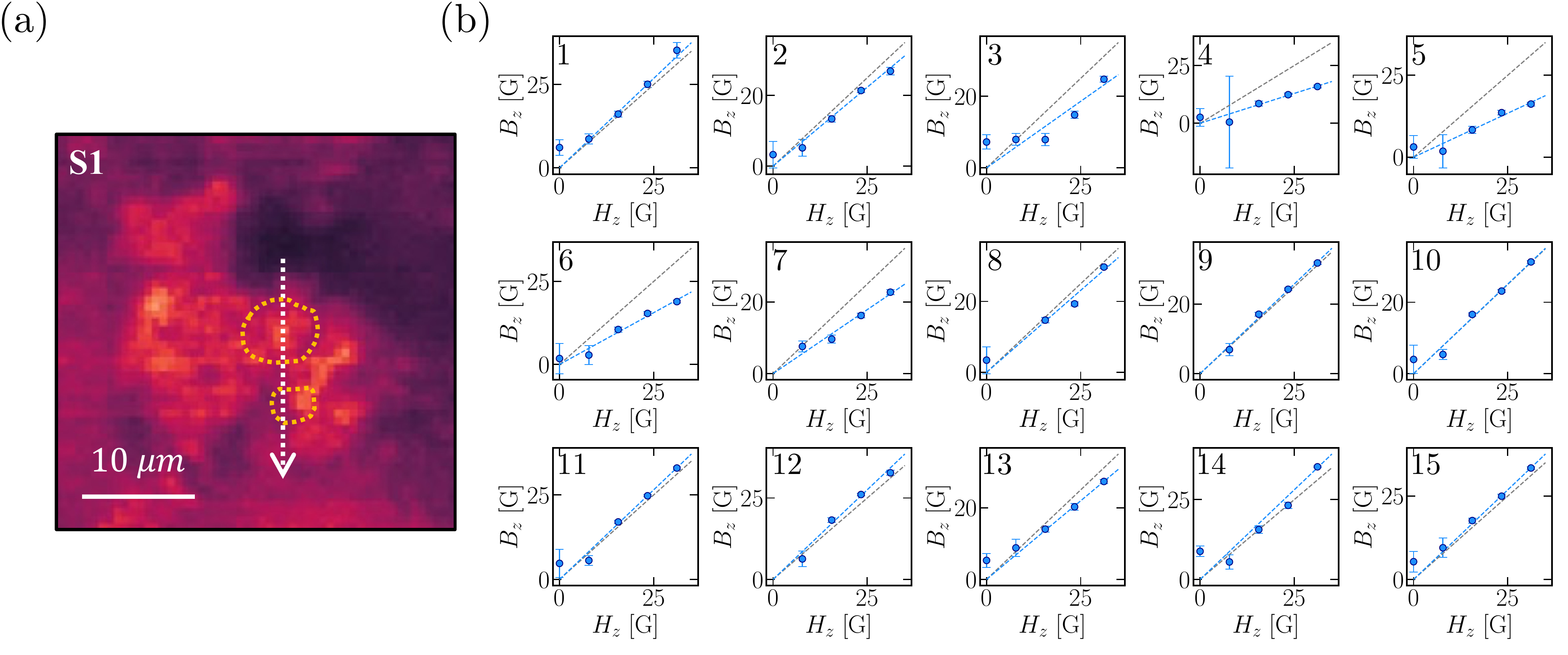}
         \caption{
           \textbf{Spatial studies of diamagnetism in S1.} 
           \textbf{(a)} We measure the local suppression in $B_z$ along a line cut in sample S1.
           The slope $s= \Delta B_z / \Delta H_z$ for this study is presented in maintext Fig.~3(g). 
           \textbf{(b)} We show plots of the extracted $B_z$ against the applied external field $H_z$ at fifteen spatial points (indexed) along the line cut. 
           Linear fits to the extracted values of $B_z$ show a clear suppression in two regions of the sample (enclosed in dotted yellow line in (a)).
           }
         \label{fig:S1-spatial-studies}
\end{figure}

\begin{figure}[h!]
         \centering
         \includegraphics[width = \textwidth]{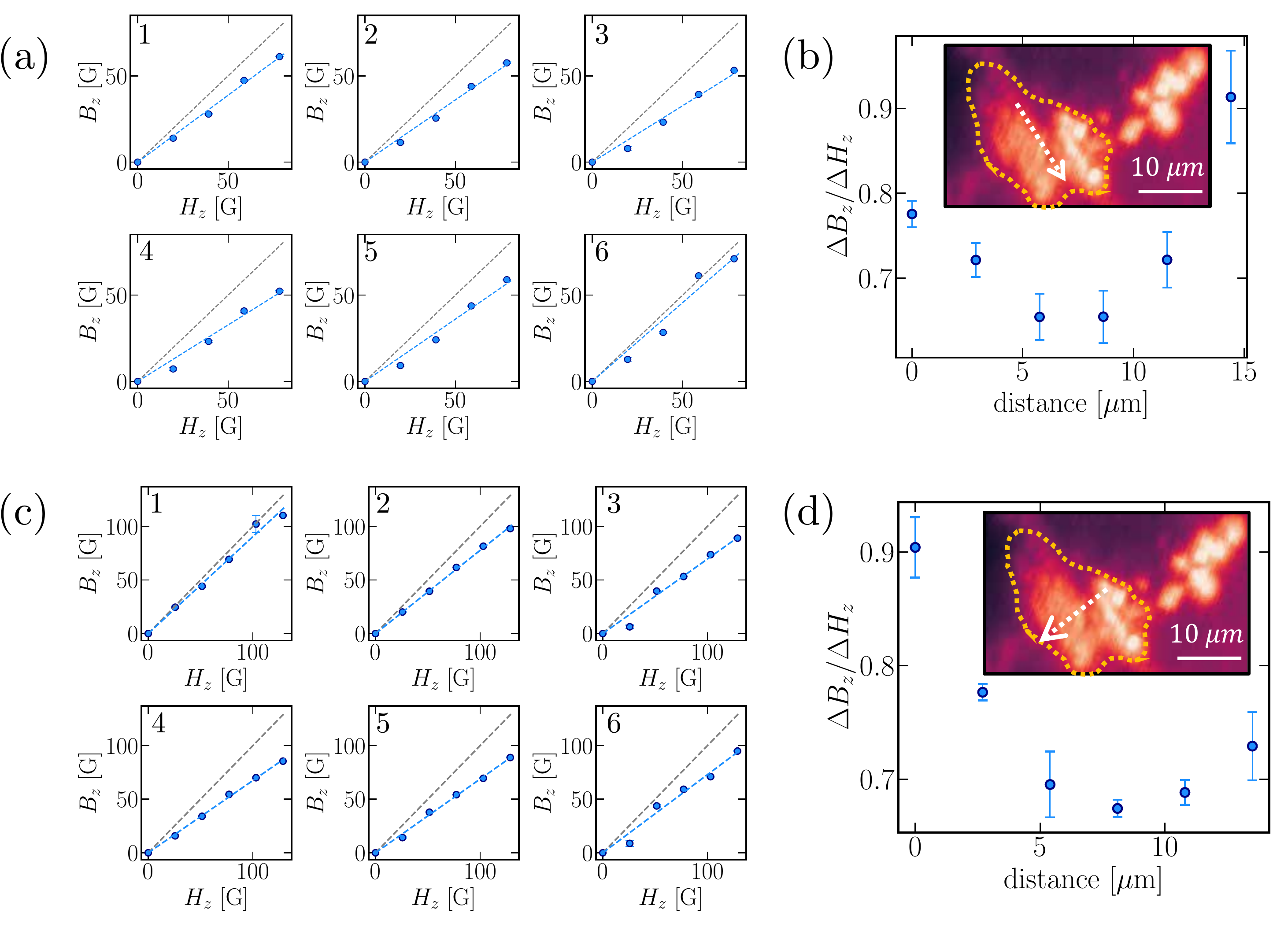}
         \caption{
           \textbf{Spatial studies of diamagnetism in S2.}
           Spatial studies in sample S2 along two orthogonal line cuts [insets in (b) and (d)] on top of the identified CeH\textsubscript{9} region (enclosed in dotted yellow). 
           \textbf{(a)} The $B_z$ values extracted from ODMR spectra are plotting against the applied external field, $H_z$, for six spatial points (indexed) along the line cut shown in inset of (b).
           \textbf{(b)} Similar to sample S1 [maintext Fig.~3(g)], we measure a spatially varying suppression in the local field suggesting the formation of $\approx10$\SI{}{\micro\meter} regions of CeH\textsubscript{9} via laser heating.
           \textbf{(c-d)} A repeat of the study for six spatial points along an orthogonal line cut (inset of (d)).
           }
         \label{fig:S2-spatial-studies}
\end{figure}

In the maintext, we present our studies of the magnetic response of sample S2 on zero field cooling and field cooling [Fig.~4].
In particular, we perform NV magnetometry and four point resistance measurements simultaneously to see concurrent transitions in electrical transport and magnetism across $T_\textrm c$.
Here, we show our studies of sample S1, where we perform NV magnetometry and resistance measurements separately [Fig.~\ref{fig:S1-suppression-studes}(d,e)].
Due to coupling between the Pt wire for MW delivery and the transport leads, accurate determination of four point resistance on S1 is only possible when the Pt wire is disconnected from the MW circuitry.
We measure a difference of $\approx9$~K between the magnetic transition [green data in Fig.~\ref{fig:S1-suppression-studes}(e)] and the resistive transition [grey data in Fig.~\ref{fig:S1-suppression-studes}(e)]. 
This difference may be due to a change in the temperature gradient between the DAC and the cold finger upon isolating the sample from the MW circuitry. 
%

\subsection{Field cooling and hysteresis in S2}

\begin{figure}[h!]
         \centering
         \vspace{-10mm}
         \includegraphics[width=\textwidth]{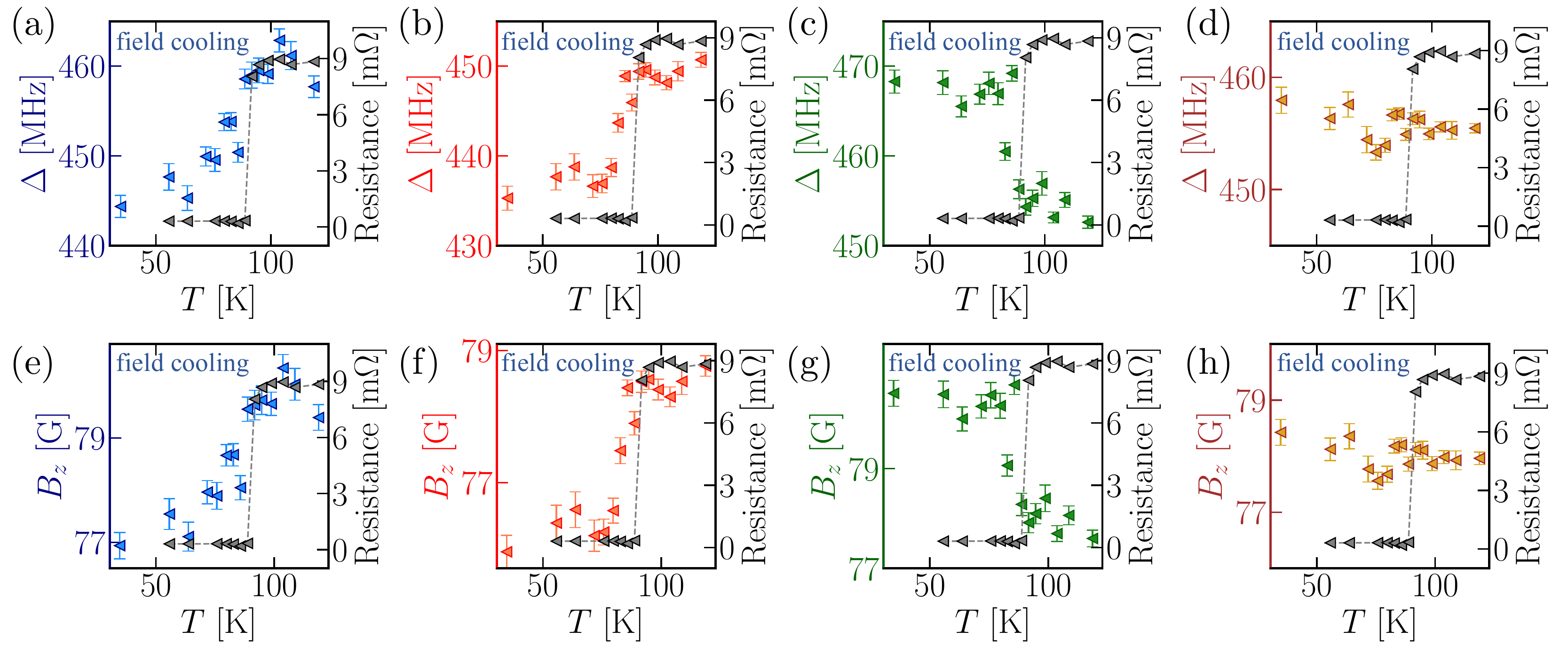}
         \caption{
           \textbf{Comparison of ODMR splitting and $B_z$ on field cooling S2 at $H_z= 79$~G.} 
           Measurements of four point electrical resistance and the ODMR splitting, $\Delta$, at four spatial locations [shown in maintext Fig.~4(b)]: \textbf{(a,b)} two points on top the synthesized CeH\textsubscript{9} (blue and orange), \textbf{(c)} one point at the edge of this region (green), and \textbf{(d)} one point away from  region (yellow).
           (a,b) On top of the CeH\textsubscript{9} region, we measure a decrease in $\Delta\approx(2\pi)\times15$~MHz as we cool below the transition point.
           (c) In contrast, at the edge of this region, we measure an increase in $\Delta\approx(2\pi)\times15$~MHz.
           (d) Away from the CeH\textsubscript{9} region we do not measure an appreciable change in $\Delta$ across the transition point (determined via simultaneous electrical resistance measurements). 
           \textbf{(e,f,g,h)} For the $\Delta$ values measured at each spatial point, we extract the magnetic field, $B_z$, based on the $2\Pi_\perp$ stress splitting (measured at the same spatial point at $H_z=0$~G after zero field cooling to $T=86$~K).
           Systematic disagreement between $B_z$ and $H_z$ for $T>T_\textrm{c}$ may stem from inaccurate determination of $2\Pi_\perp$ or change in the value of $2\Pi_\perp$ with temperature.
           }
           \label{FC-splitting-Bz-comparison}
\end{figure}

In the maintext, we show four point electrical resistance measurements and the change in the local field, $\delta B_z$, at four spatial locations in sample S2 upon field cooling at $H_z=79$~G [Fig.~4(e)]. 
We determine $\delta B_z$ at each spatial point by subtracting the mean $B_z$ measured at the corresponding location in the normal state (i.e., for all temperature points $T>91$~K).
Here, we show the ODMR splitting, $\Delta$, and the extracted local field, $B_z$, measured at these four spatial locations for the same experiment [Fig.~\ref{FC-splitting-Bz-comparison}].
On top of the synthesized CeH\textsubscript{9} region, we measure a decrease in $\Delta$ by $\sim(2\pi)\times15$~MHz and in $B_z$ by $\sim2$~G upon cooling below the transition point, $T_\textrm{c}$. 
At the edge of this region, we measure an increase in $\Delta$ by $\sim(2\pi)\times15$~MHz and in $B_z$ by $\sim2$~G across the transition point. 
Finally, away from this region we do not measure an appreciable change in $\Delta$ or $B_z$ across the transition.
At each spatial location, we extract $B_z$ from $\Delta$ by using the $2\Pi_\perp$ stress splitting (measured at the respective location at $H_z=0$~G after zero field cooling to $T=86$~K).
In the normal state ($T>T_\textrm{c}$), we find systematic differences between the applied field, $H_z\approx79$~G, and the extracted values of $B_z$.
These may be due to changes in stress splitting, $2\Pi_\perp$, with temperature.
In addition, they can also stem from any inaccuracy in the determination of $2\Pi_\perp$ due to spatial drift in the sample.
Despite these differences, we observe a clear signal in the ODMR splitting, $\Delta$, suggesting flux expulsion within the CeH\textsubscript{9} region.

\begin{figure}[h!]
         \centering
         \vspace{-10mm}
         \includegraphics[width = 0.75\textwidth]{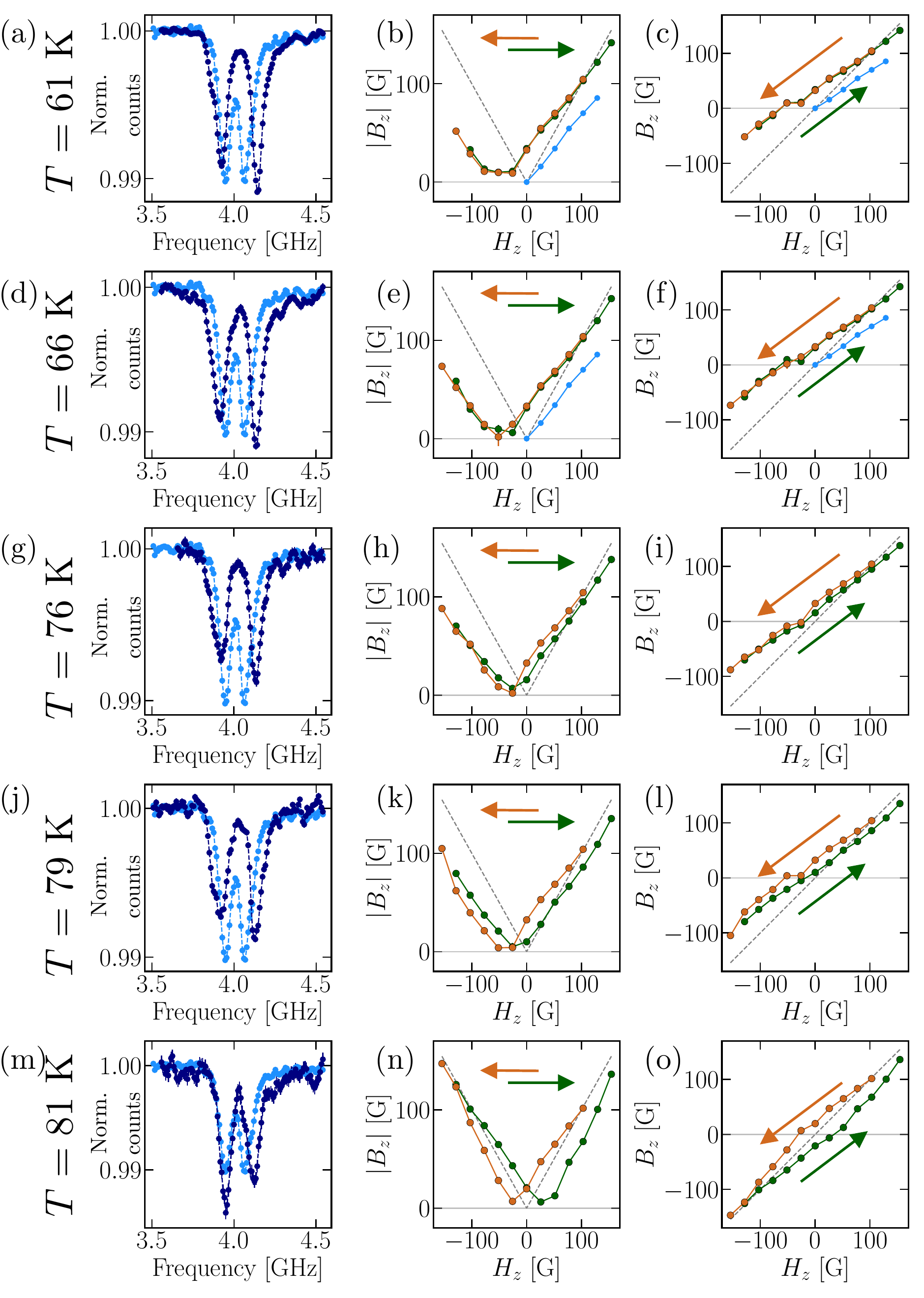}
         \caption{
           \textbf{Study of flux trapping in sample S2 (full dataset).} 
           %
           We field cool sample S2 at $H_z=103$~G to several temperatures (indicated on the left) and ramp the magnitude and direction of $H_z$ [experimental sequence in Fig.~5(f)].
           \textbf{(a,d,g,j,m)} 
           Light blue data show the ODMR spectrum measured at $H_z=0$ on zero field cooling to $T=81$~K.
           On ramping to $H_z=0$ after field cooling to several temperatures, we observe a markedly higher splitting in the ODMR spectra (dark blue data) suggesting the presence of a remnant flux trapped field. 
           The strength of the flux trapped field (i.e., $B_z$ values at $H_z=0$) decreases with increasing temperature. 
           \textbf{(b,e,h,k,n)} 
           In order to extract the local $|B_z|$ field from the splitting ($\Delta$), we use the $2\Pi_{\perp}$ splitting measured at $H_z= 0$ zero field cooling to $T=81$~K [light blue data in (a,d,g,j,m)].
           On switching the direction of the external field ($H_z<0$), we measure a minimum in $|B_z|$.
           Since the local field measured at the NV location is the sum of the flux trapped field and $H_z$, we see an increase in the splitting on further increasing the magnitude of the external field in the negative direction.
           \textbf{(c,f,i,l,o)} Based on the continuity of the $|B_z|$ across the minimum, we assign a direction (sign) to the local field to get the final hysteresis plots.
           The local suppression in $B_z$ at the same spatial location on ZFC [blue data included in (b,c,e,f)] shows the slope.
           All measurements are performed at the spatial location shown by the white square in maintext Fig.~5(c) [also shown in inset in Fig.~\ref{fig:Hz-hysteresis-FC-ZFC-comparison}(a)].
           }
         \label{fig:Hz-hysteresis-FC}
\end{figure}

For sample S2, we also study flux trapping on field cooling (at $H_z=103$~G) to several temperatures $T<T_\textrm{c}$~[Fig.\ref{fig:Hz-hysteresis-FC}]. 
We measure the ODMR splitting $\Delta$ as a function of $H_z$. 
On ramping down to $H_z=0$, we clearly measure $\Delta > 2\Pi_{\perp}$ [Fig.~\ref{fig:Hz-hysteresis-FC}(a,d,g,j,m)] suggesting the presence of a remnant magnetic field at the location of the NV center due to flux trapping.
By sweeping both magnitude and direction of the external field, we characterize the hysteresis of the trapped flux.
In particular, we observe a minimum in the ODMR splitting ($\Delta\sim2\Pi_{\perp}$) for $H_z<0$ where the external field is anti-aligned to the flux trapped field. 
On ramping $H_z$ to larger negative values, the splitting increases because the external field dominates the flux trapped field [Fig.~\ref{fig:Hz-hysteresis-FC}(b,e,h,k,n)].
Although, we can only extract the absolute value of the local field, $|B_z|$, from $\Delta$, we are able to clearly designate a sign (direction) to $B_z$ based on the continuity in the splitting profile [Fig.~\ref{fig:Hz-hysteresis-FC}(c,f,i,l,o)]. 

We study the effect of field cooling at several spatial locations in sample S2 [Fig.~\ref{fig:S2-FC-ZFC-other-studies}]. 
We find similar signatures of flux trapping at two spatial points on top of the CeH\textsubscript{9} region [Fig.~\ref{fig:S2-FC-ZFC-other-studies}(a,b,d,e)].
At a spatial point away from the sample, we do not observe a magnetic response on field cooling or on zero field cooling [Fig.~\ref{fig:S2-FC-ZFC-other-studies}(c,f)].

\subsection{Flux penetration in S2 on ZFC and FC}

\begin{figure}[h!]
         \centering
         \includegraphics[width = \textwidth]{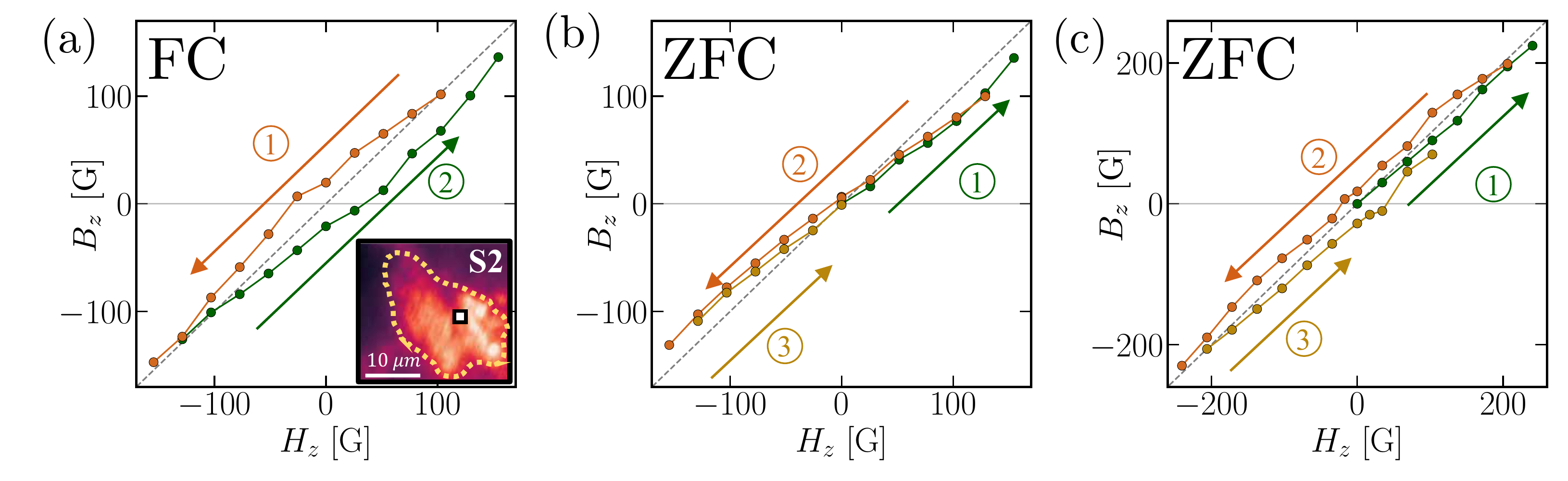}
         \caption{
           \textbf{Flux trapping in S2.}
           We compare the response of the sample on sweeping $H_z$ after field cooling and zero field cooling at $T=81$~K. 
           \textbf{(a)} On field cooling at $H_z=103$~G we measure a trapped flux. 
           On ramping to large fields in the direction opposite to this trapped flux ($H_z=-154$~G), we measure an inversion of the flux trapped field. 
           \textbf{(b)} In contrast, the sample is robust to flux penetration on zero field cooling. 
           Specifically, on ramping the external field to $|H_z|=154$~G in both the positive and negative directions, we continue to observe a local suppression in $B_z$ without any appreciable flux trapping.
           \textbf{(c)} After zero field cooling, the sample exhibits flux penetration only on ramping the external field to much larger magnitudes ($|H_z|\sim240$~G). 
           Here, we measure a response similar to the case of field cooling.
           Specifically, on ramping to a field $H_z=240$~G we find a flux trapped field $B_z=18$~G (measured after returning to $H_z=0$~G). 
           Similarly, after ramping to a field $H_z=-240$~G, the direction of the trapped flux changes (i.e., we measure $B_z=-28$~G on returning back to $H_z=0$~G).
           All measurements are performed at the spatial location shown by the white square in the inset of (a).
           }
         \label{fig:Hz-hysteresis-FC-ZFC-comparison}
\end{figure}

\begin{figure}[h!]
         \centering
         \includegraphics[width = \textwidth]{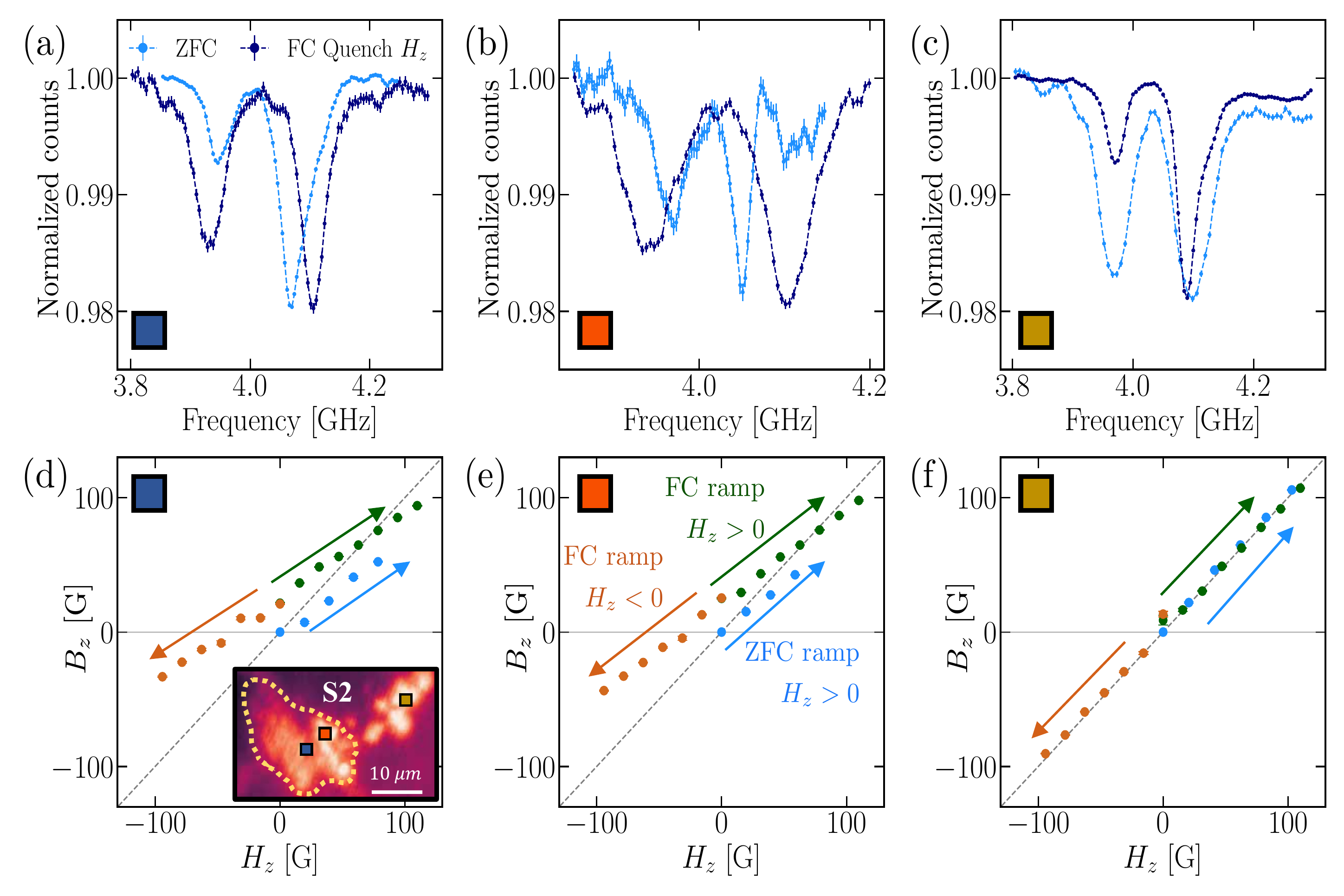}
         \caption{
           \textbf{Spatial studies of flux trapping in S2.} 
           We compare the effects of zero field cooling (ZFC) and field cooling (FC) at several spatial points in sample S2: two locations [blue and red points in inset of (d)] on top of the CeH\textsubscript{9} region and one location [yellow point in inset of (d)] away from this region.
           \textbf{(a-c)} We field cool the sample at $H_z=79$~G to a temperature $T=36$~K and measure the ODMR spectra (dark blue) after quenching the external field to $H_z=0$~G.
           Light blue data show the ODMR spectra measured at the respective spatial locations upon zero field cooling to temperatures (a,b) $T=36$~K, and (c) $T=86$~K.
           We see markedly higher splittings on top of the CeH\textsubscript{9} region (a,b) upon field cooling indicating the presence of a flux trapped field.
           Away from the CeH\textsubscript{9} region (c), we do not observe a significant difference between ZFC and FC data.
           In particular, $\Delta$ for the ZFC spectrum (measured at $T=86$~K) is larger by $7$~MHz; this difference is likely due to temperature dependence of the stress splitting, $2\Pi_{\perp}$, at this spatial point.
           \textbf{(d-f)} Following the field quench after FC, we determine the local $B_z$ field at the three spatial locations on ramping $H_z$ to positive values (green data points). 
           We make the same measurements on ramping $H_z$ to negative values (orange data points).
           We also measure $B_z$ as a function of $H_z$ after zero field cooling (blue data points).
           (d,e) On top of the CeH\textsubscript{9} region, our measurements indicate the presence of a flux trapped field upon FC. 
           This contrasts with a local suppression of $H_z$ measured at the same locations upon ZFC.
           In particular, the slope $s=\Delta B_z/\Delta H_z$ on field cooling is in quantitative agreement with the slope on zero field cooling.
           (f) Away from CeH\textsubscript{9} region, we see no significant difference in response on ramping $H_z$ after FC and ZFC.
           To extract $B_z$ at each point, we use the $2\Pi_{\perp}$ splitting measured upon zero field cooling at the respective spatial location [light blue spectra in (a-c)]. 
           }
         \label{fig:S2-FC-ZFC-other-studies}
\end{figure}

In our study of hysteresis of the trapped flux [Fig.~\ref{fig:Hz-hysteresis-FC}], we measure an overall inversion of the flux trapped field at temperatures closer to $T_\textrm c$. 
Specifically, after ramping to a field $H_z\sim-154$~G at $T=81$~K, we find the flux trapped field is inverted (i.e., $B_z<0$ on returning to $H_z=0$~G) [Fig.~\ref{fig:Hz-hysteresis-FC-ZFC-comparison}(a)].
Here, we conduct a complementary study of the sample's response to $H_z$-sweeps after zero field cooling wherein there should be no flux initially trapped. 
Specifically, we zero field cool to $T=81$~K and study the response on ramping the magnitude and direction of $H_z$.
Interestingly, we find that Meissner suppression dominates for fields $|H_z|\lesssim154$~G and there is no appreciable penetration of flux lines into the sample [Fig.~\ref{fig:Hz-hysteresis-FC-ZFC-comparison}(b)].  
However, on ramping to larger fields ($|H_z|\sim240$~G) after zero field cooling to $T=81$~K [Fig.~\ref{fig:Hz-hysteresis-FC-ZFC-comparison}(c)], we observe the penetration and trapping of external flux into the sample (i.e., we measure $B_z\neq0$ on ramping back to $H_z=0$).
Our observations suggest that the penetration of external flux depends on the amount of flux initially trapped in the sample.

\subsection{Hysteresis on $T$-sweeps in S2}

\begin{figure}[h!]
         \centering
         \includegraphics[width = 0.8\textwidth]{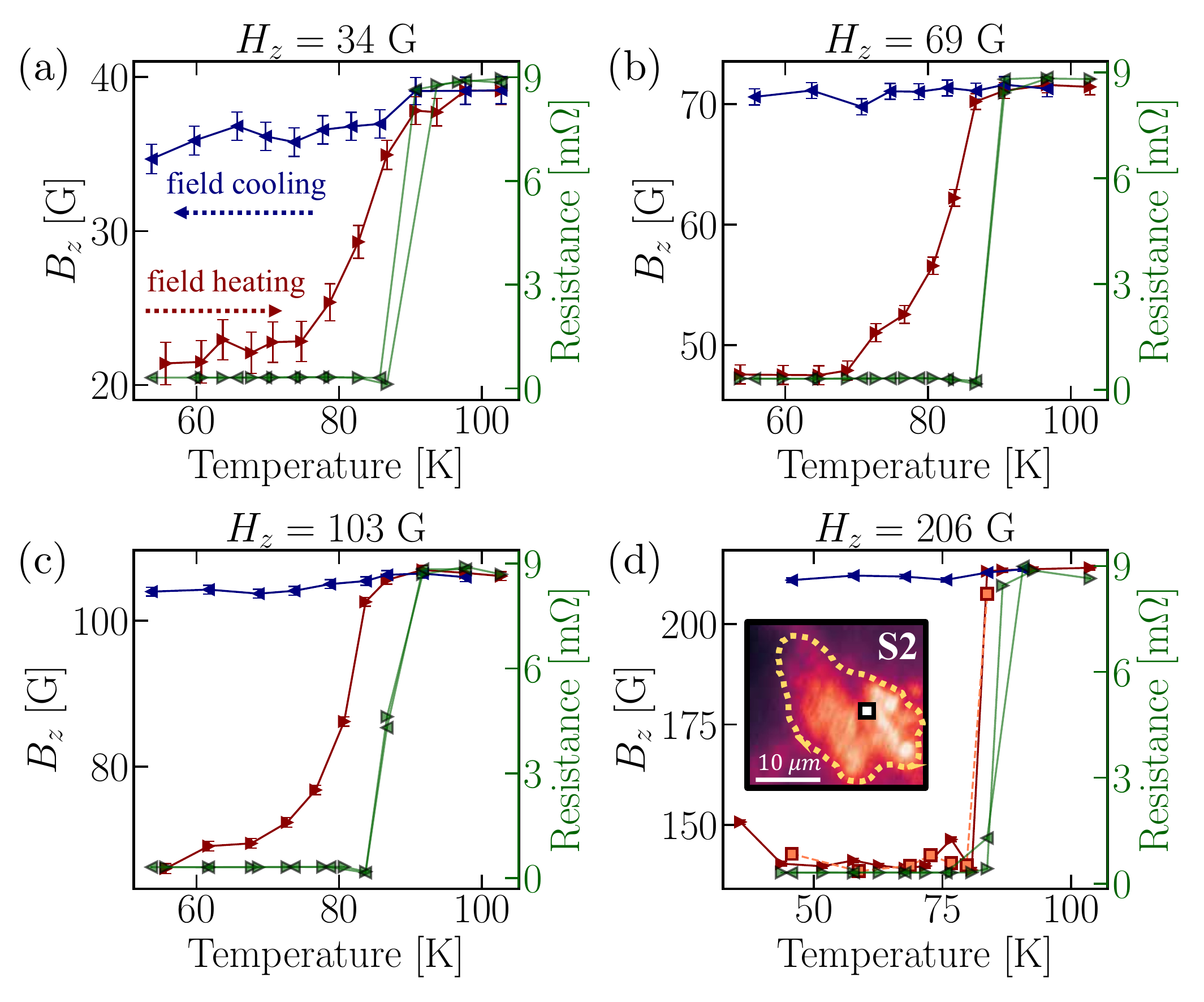}
         \caption{
           \textbf{Hysteresis on $T$-sweeps in S2.}
            To study hysteresis on sweeping $T$, we first zero field cool the sample and apply an external field $H_z$ [maintext Fig.~5~(d)]. 
            Then, fixing $H_z$, we heat the sample across the transition (field heating shown in red). 
            Finally, we cool the sample below $T_\textrm c$ without changing $H_z$ (field cooling shown in blue). 
            We perform simultaneous electrical resistance (shown in green) and ODMR measurements.
            Four point resistance measured on field heating (field cooling) are indicated by markers pointing right (left).
            All measurements are made at the spatial location shown by the white square in inset of (c). 
            We use the $2\Pi_{\perp}$ stress parameter measured at $H_z=0$ and $T=81$~K to extract the local $B_z$ field for all data points.
            \textbf{(a-c)} We observe concurrent transitions in magnetism and electrical resistance.
            In addition, we see a suppression of $T_\textrm c$ with the increase in the magnitude of $H_z$.
            \textbf{(d)} Surprisingly at high fields ($H_z=206$~G), we see a clear sharpening of the magnetic transition. 
            To confirm the signal we repeat the measurement on field heating near the transition region (orange squares in (d)). 
           }
         \label{fig:T-hysteresis}
\end{figure}

We include the full data set of our study of hysteresis of diamagnetism in S2 on sweeping sample temperature, $T$, at a fixed value of the external field, $H_z$. 
We observe a characteristic loss of suppression on field heating across $T_\textrm c$ concurrent with a jump in the four point resistance. 
We observe a surprising sharpening of the magnetic transition at the highest field [$H_z=206$~G in Fig.~\ref{fig:T-hysteresis}(d)].
A repeated measurement along the transition profile for field heating at $H_z=206$~G  [orange squares in Fig.~\ref{fig:T-hysteresis}(d)] further verifies this observation. 

\subsection{A comment on errors}

In our calculation of $B_z$, we perform error propagation of fitting error for gaussian fits to the ODMR spectra. 
These errorbars are included in all our plots.
We note that this may not necessarily reflect the true error in our measurements.
In this section, we list the different sources of error. 

\begin{figure}[h!]
         \centering
         \includegraphics[width = 0.8\textwidth]{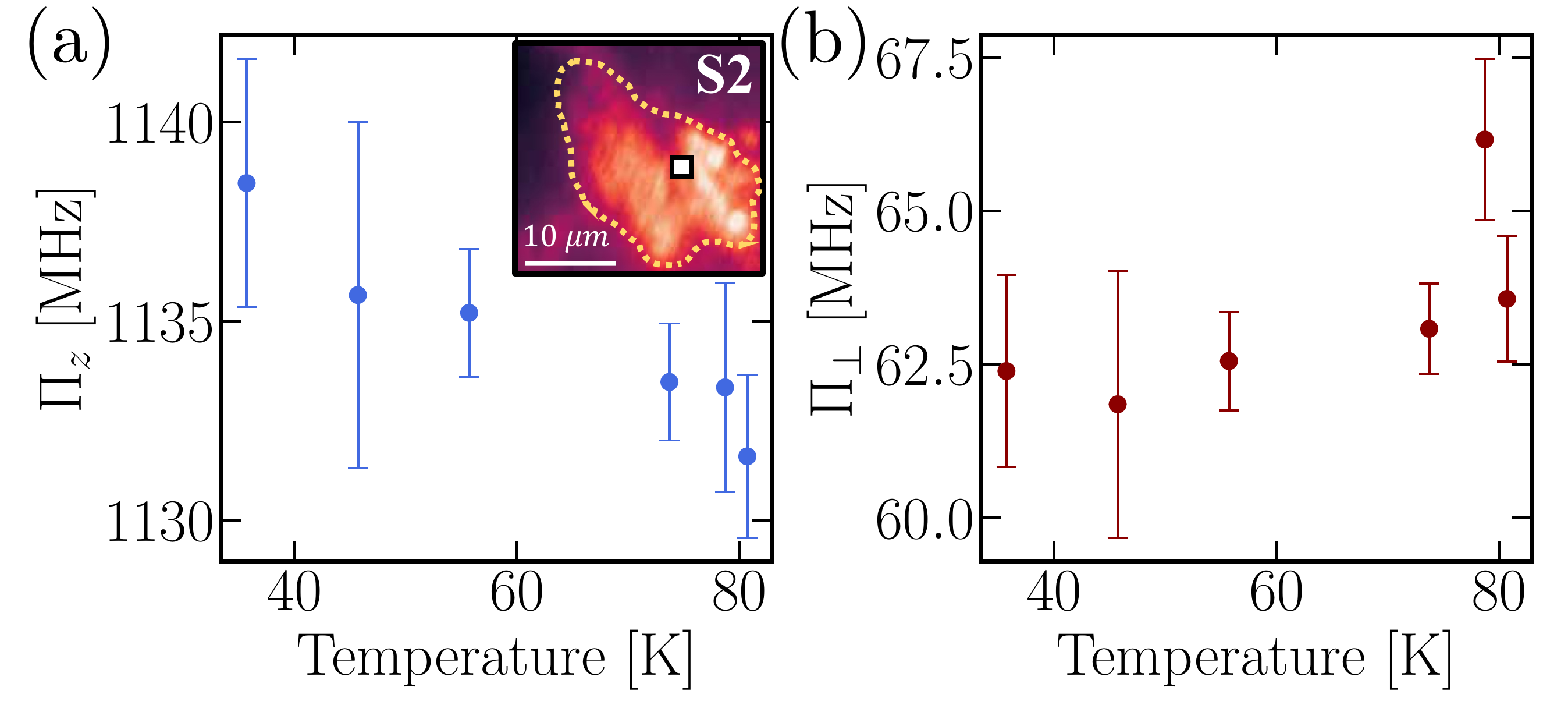}
         \caption{\textbf{Change in stress parameters with temperature in S2.}
         We directly extract the \textbf{(a)} shift, $\Pi_z$, and \textbf{(b)} splitting, $\Pi_{\perp}$, from ODMR spectra measured at several temperatures at the same spatial location (white point in inset of (a)). 
         To extract $\Pi_z$, we subtract the temperature dependent zero-field splitting, $D_{\textrm{gs}}$, from the center frequency of the ODMR resonances~\cite{chen2011temperature}.
         We find that $\Pi_z$ increases with decreasing temperature consistent with an expected increase in sample pressure~\cite{gavriliuk2009miniature}.
         For this sample, we observe a comparatively smaller change in the $\Pi_{\perp}$ stress parameter (within $\sim(2\pi)\times8$~MHz). 
         }
         \label{fig:DE-S2-temperature-change}
\end{figure}

\begin{itemize}
    \item A change in temperature induces volumetric changes in the DAC body~\cite{gavriliuk2009miniature}; this results in different stress parameters $\{\Pi_z,\Pi_{\perp}\}$ for different temperatures [Fig.~\ref{fig:DE-S2-temperature-change}].
    The extraction of $B_z$ at a specific temperature relies on the accurate determination of the stress splitting ($2\Pi_{\perp}$) at the same temperature.
    Most of these studies are performed on sample S2, where we find that  temperature dependent changes in the $\Pi_{\perp}$ splitting are within $\approx(2\pi)\times8$~MHz [Fig.~\ref{fig:DE-S2-temperature-change}(b)].
    However, it was not feasible to determine the temperature dependence of $2\Pi_{\perp}$ at all relevant spatial locations.
    For each spatial location in Fig.~4(e-h), we extract $B_z$ at all temperatures points by using the $2\Pi_{\perp}$ splitting measured at a single temperature point ($T=86$~K) at the respective spatial location.
    The systematic disagreement between $B_z$ and $H_z$ for $T>T_\textrm{c}$ for this data set is likely due to minor changes in the $2\Pi_{\perp}$ parameter across the temperature sweep.
    In particular, an error of $(2\pi)\times8$~MHz in the $\Pi_{\perp}$ parameter translates to an error of $1.7$~G in the determination of $B_z$ for this study.
    Owing to the same reason, we find similarly small systematic disagreements for data shown in Fig.~\ref{fig:T-hysteresis}.
    This issue may be mitigated by switching to widefield microscopy~\cite{hsieh2019imaging,lesik2019magnetic,scholten2021widefield}, where one can determine the temperature dependence of $\{\Pi_z,\Pi_{\perp}\}$ across the entire culet more efficiently.
    
    \item We note that laser heating at high pressure ($>100$~GPa) results in spatial variation of $2\Pi_{\perp}$ values across the sample region.
    Despite careful monitoring, small variations in the measurements may be induced by spatial drifts in the sample during data acquisition. 
    In almost all cases, the extracted signal magnitudes outweigh errors introduced in this process.
    Image registration in widefield datasets can correct for these errors in future~\cite{van2014scikit}. 
    
    \item ODMR spectra in the laser heated regions sample the stress gradient over the depth of a few microns [Fig.~\ref{fig:laser-heating-NV-creation}(c)]. 
    Owing to this, we do not observe clear gaussian lineshapes at most spatial locations in these regions. 
    To simplify our analysis, we restrict our studies to spatial locations where the ODMR lineshapes are approximately gaussian. 

\end{itemize}
\subsection{A comparison with SQUID magnetometry at high pressure}

Magnetic measurements of high pressure superhydrides typically rely on SQUID magnetometry where dipole sensitivity is on the order of $\approx10^{-8}$~emu~\cite{drozdov2015conventional,Minkov_2022,eremets2022high}.
Although expected signals from superhydride samples beat this threshold, the subtraction of parasitic pick up from a paramagnetic backgrounds constitutes an enormous challenge.
In addition, by averaging over the entire DAC geometry, such global probes discard local spatial information of superconducting samples.
We demonstrate continuous wave ODMR spectroscopy, with typical magnetic sensitivity $\sim~35~\SI{}{\micro\tesla/\sqrt{\hertz}}$ at room temperature.
Following previous work, this corresponds to a dipole sensitivity on the order $\approx10^{-11}$~\SI{}{emu/\sqrt{\hertz}} at room temperatures ($\approx 10^{-10}$~\SI{}{emu/\sqrt{\hertz}} at cryogenic temperatures)~\cite{hsieh2019imaging}.
In this case, a key advantage is the proximity of the sensor to the target material and the ability to measure magnetism with diffraction limited sub-micron spatial resolution.

\bibliography{supplementary.bib}

\newpage

\newpage

\newpage

\newpage

\newpage

\newpage

\newpage

\newpage

\newpage

\newpage

\newpage

\newpage

\newpage

\newpage

\renewcommand{\arraystretch}{2}